\documentclass[longauth]{aa}

\usepackage{amsmath,amssymb,amsfonts}
\usepackage{graphicx}
\usepackage{txfonts}
\usepackage{enumerate} 
\usepackage{colordvi} 
\usepackage{bm}
\usepackage{epsfig}
\usepackage{hyperref}
\usepackage[dvipsnames]{xcolor}
\usepackage{color, colortbl}
\usepackage[normalem]{ulem}
\usepackage[T1]{fontenc} 
\usepackage{braket}
\usepackage{euclid}
\usepackage{placeins}
\usepackage{multirow}
\usepackage{tabularx}
\usepackage{comment}
\usepackage[normalem]{ulem}

\usepackage{natbib}
\usepackage{tabularx}

\usepackage{colortbl} 
\usepackage{bm}

\usepackage{fontawesome}
\usepackage{textcomp}
\usepackage{hyperref}
\hypersetup{
    colorlinks=true,
    linkcolor=blue,
    filecolor=magenta,      
    urlcolor=blue,
    citecolor=blue
}

\usepackage[nameinlink,noabbrev]{cleveref}

\Crefname{section}{Section}{Sections}
\crefname{section}{Sect.}{Sects.}
\Crefname{figure}{Figure}{Figures}
\crefname{figure}{Fig.}{Figs.}
\Crefname{table}{Table}{Tables}
\crefname{table}{Table}{Tables}
\Crefname{equation}{Equation}{Equations}
\crefname{equation}{Eq.}{Eqs.}

\usepackage{xcolor}
\usepackage[normalem]{ulem} 

\makeatletter
\renewcommand*\aa@pageof{, page \thepage{} of \pageref*{LastPage}}
\makeatother

\usepackage[utf8]{inputenc}
\usepackage{booktabs}  
\usepackage{siunitx}   
\usepackage[switch, modulo]{lineno}
\renewcommand{\linenumbers}[0]{}

\makeatletter
\newcommand{\deceasedsym}{\textsuperscript{\@fnsymbol{3}}}
\makeatother

\newcommand{\de}{\mathrm{d}}
\newcommand{\pd}{\partial}
\newcommand{\sfr}{\ensuremath{\mathrm{SFR}}}
\newcommand{\rsfr}{\ensuremath{\rho_\mathrm{SFR}}}
\newcommand{\rsfrd}{\ensuremath{\rho_\mathrm{SFRD}}}
\newcommand{\brsfr}{\ensuremath{\langle b \rsfr\rangle}}
\newcommand{\brsfrd}{\ensuremath{\langle b \rsfrd\rangle}}
\DeclareMathOperator{\cov}{Cov}

\let\Delta\varDelta
\let\Theta\varTheta
\let\Xi\varXi
\let\Pi\varPi
\let\Sigma\varSigma
\let\Upsilon\varUpsilon
\let\Phi\varPhi
\let\Psi\varPsi
\let\Omega\varOmega

\defcitealias{maniyar2021simple}{M21}

\begin{document}

\title{\Euclid: Inferring star-formation history via cross-correlations of photometric clustering and shear with the cosmic infrared background\thanks{This paper is published on behalf of the Euclid Consortium.}}    
\newcommand{\orcid}[1]{} 
\author{J.~Han\orcid{0009-0008-5180-0039}\thanks{\email{Jiakang.han@unito.it}}\inst{\ref{aff1},\ref{aff2},\ref{aff3}}
\and S.~Camera\orcid{0000-0003-3399-3574}\inst{\ref{aff1},\ref{aff2},\ref{aff3}}
\and M.~Migliaccio\inst{\ref{aff4},\ref{aff5}}
\and G.~Fabbian\orcid{0000-0002-3255-4695}\inst{\ref{aff6}}
\and F.~Pace\orcid{0000-0001-8039-0480}\inst{\ref{aff1},\ref{aff2},\ref{aff3}}
\and B.~Jego\orcid{0009-0006-6399-7858}\inst{\ref{aff7}}
\and B.~Altieri\orcid{0000-0003-3936-0284}\inst{\ref{aff8}}
\and S.~Andreon\orcid{0000-0002-2041-8784}\inst{\ref{aff9}}
\and N.~Auricchio\orcid{0000-0003-4444-8651}\inst{\ref{aff10}}
\and C.~Baccigalupi\orcid{0000-0002-8211-1630}\inst{\ref{aff11},\ref{aff12},\ref{aff13},\ref{aff14}}
\and M.~Baldi\orcid{0000-0003-4145-1943}\inst{\ref{aff15},\ref{aff10},\ref{aff16}}
\and S.~Bardelli\orcid{0000-0002-8900-0298}\inst{\ref{aff10}}
\and P.~Battaglia\orcid{0000-0002-7337-5909}\inst{\ref{aff10}}
\and A.~Biviano\orcid{0000-0002-0857-0732}\inst{\ref{aff12},\ref{aff11}}
\and E.~Branchini\orcid{0000-0002-0808-6908}\inst{\ref{aff17},\ref{aff18},\ref{aff9}}
\and M.~Brescia\orcid{0000-0001-9506-5680}\inst{\ref{aff19},\ref{aff20}}
\and G.~Ca\~nas-Herrera\orcid{0000-0003-2796-2149}\inst{\ref{aff21}}
\and V.~Capobianco\orcid{0000-0002-3309-7692}\inst{\ref{aff3}}
\and C.~Carbone\orcid{0000-0003-0125-3563}\inst{\ref{aff22}}
\and V.~F.~Cardone\inst{\ref{aff23},\ref{aff24}}
\and J.~Carretero\orcid{0000-0002-3130-0204}\inst{\ref{aff25},\ref{aff26}}
\and M.~Castellano\orcid{0000-0001-9875-8263}\inst{\ref{aff23}}
\and G.~Castignani\orcid{0000-0001-6831-0687}\inst{\ref{aff10}}
\and S.~Cavuoti\orcid{0000-0002-3787-4196}\inst{\ref{aff20},\ref{aff27}}
\and K.~C.~Chambers\orcid{0000-0001-6965-7789}\inst{\ref{aff28}}
\and A.~Cimatti\inst{\ref{aff29}}
\and C.~Colodro-Conde\inst{\ref{aff30}}
\and G.~Congedo\orcid{0000-0003-2508-0046}\inst{\ref{aff31}}
\and L.~Conversi\orcid{0000-0002-6710-8476}\inst{\ref{aff32},\ref{aff8}}
\and Y.~Copin\orcid{0000-0002-5317-7518}\inst{\ref{aff33}}
\and F.~Courbin\orcid{0000-0003-0758-6510}\inst{\ref{aff34},\ref{aff35},\ref{aff36}}
\and H.~M.~Courtois\orcid{0000-0003-0509-1776}\inst{\ref{aff37}}
\and M.~Cropper\orcid{0000-0003-4571-9468}\inst{\ref{aff38}}
\and H.~Degaudenzi\orcid{0000-0002-5887-6799}\inst{\ref{aff39}}
\and S.~de~la~Torre\inst{\ref{aff40}}
\and G.~De~Lucia\orcid{0000-0002-6220-9104}\inst{\ref{aff12}}
\and H.~Dole\orcid{0000-0002-9767-3839}\inst{\ref{aff6}}
\and F.~Dubath\orcid{0000-0002-6533-2810}\inst{\ref{aff39}}
\and X.~Dupac\inst{\ref{aff8}}
\and S.~Dusini\orcid{0000-0002-1128-0664}\inst{\ref{aff41}}
\and S.~Escoffier\orcid{0000-0002-2847-7498}\inst{\ref{aff42}}
\and M.~Farina\orcid{0000-0002-3089-7846}\inst{\ref{aff43}}
\and R.~Farinelli\inst{\ref{aff10}}
\and F.~Faustini\orcid{0000-0001-6274-5145}\inst{\ref{aff23}}
\and S.~Ferriol\inst{\ref{aff33}}
\and F.~Finelli\orcid{0000-0002-6694-3269}\inst{\ref{aff10},\ref{aff44}}
\and P.~Fosalba\orcid{0000-0002-1510-5214}\inst{\ref{aff45},\ref{aff46}}
\and S.~Fotopoulou\orcid{0000-0002-9686-254X}\inst{\ref{aff47}}
\and N.~Fourmanoit\orcid{0009-0005-6816-6925}\inst{\ref{aff42}}
\and M.~Frailis\orcid{0000-0002-7400-2135}\inst{\ref{aff12}}
\and E.~Franceschi\orcid{0000-0002-0585-6591}\inst{\ref{aff10}}
\and M.~Fumana\orcid{0000-0001-6787-5950}\inst{\ref{aff22}}
\and S.~Galeotta\orcid{0000-0002-3748-5115}\inst{\ref{aff12}}
\and K.~George\orcid{0000-0002-1734-8455}\inst{\ref{aff48}}
\and B.~Gillis\orcid{0000-0002-4478-1270}\inst{\ref{aff31}}
\and C.~Giocoli\orcid{0000-0002-9590-7961}\inst{\ref{aff10},\ref{aff16}}
\and J.~Gracia-Carpio\orcid{0000-0003-4689-3134}\inst{\ref{aff49}}
\and A.~Grazian\orcid{0000-0002-5688-0663}\inst{\ref{aff50}}
\and F.~Grupp\inst{\ref{aff49},\ref{aff51}}
\and S.~V.~H.~Haugan\orcid{0000-0001-9648-7260}\inst{\ref{aff52}}
\and W.~Holmes\orcid{0009-0007-8554-4646}\inst{\ref{aff53}}
\and F.~Hormuth\inst{\ref{aff54}}
\and A.~Hornstrup\orcid{0000-0002-3363-0936}\inst{\ref{aff55},\ref{aff56}}
\and K.~Jahnke\orcid{0000-0003-3804-2137}\inst{\ref{aff57}}
\and M.~Jhabvala\inst{\ref{aff58}}
\and B.~Joachimi\orcid{0000-0001-7494-1303}\inst{\ref{aff59}}
\and S.~Kermiche\orcid{0000-0002-0302-5735}\inst{\ref{aff42}}
\and A.~Kiessling\orcid{0000-0002-2590-1273}\inst{\ref{aff53}}
\and B.~Kubik\orcid{0009-0006-5823-4880}\inst{\ref{aff33}}
\and M.~Kunz\orcid{0000-0002-3052-7394}\inst{\ref{aff60}}
\and H.~Kurki-Suonio\orcid{0000-0002-4618-3063}\inst{\ref{aff61},\ref{aff62}}
\and A.~M.~C.~Le~Brun\orcid{0000-0002-0936-4594}\inst{\ref{aff63}}
\and S.~Ligori\orcid{0000-0003-4172-4606}\inst{\ref{aff3}}
\and P.~B.~Lilje\orcid{0000-0003-4324-7794}\inst{\ref{aff52}}
\and V.~Lindholm\orcid{0000-0003-2317-5471}\inst{\ref{aff61},\ref{aff62}}
\and I.~Lloro\orcid{0000-0001-5966-1434}\inst{\ref{aff64}}
\and M.~Magliocchetti\orcid{0000-0001-9158-4838}\inst{\ref{aff43}}
\and G.~Mainetti\orcid{0000-0003-2384-2377}\inst{\ref{aff65}}
\and O.~Mansutti\orcid{0000-0001-5758-4658}\inst{\ref{aff12}}
\and O.~Marggraf\orcid{0000-0001-7242-3852}\inst{\ref{aff66}}
\and M.~Martinelli\orcid{0000-0002-6943-7732}\inst{\ref{aff23},\ref{aff24}}
\and N.~Martinet\orcid{0000-0003-2786-7790}\inst{\ref{aff40}}
\and F.~Marulli\orcid{0000-0002-8850-0303}\inst{\ref{aff67},\ref{aff10},\ref{aff16}}
\and R.~J.~Massey\orcid{0000-0002-6085-3780}\inst{\ref{aff68}}
\and E.~Medinaceli\orcid{0000-0002-4040-7783}\inst{\ref{aff10}}
\and S.~Mei\orcid{0000-0002-2849-559X}\inst{\ref{aff69},\ref{aff70}}
\and M.~Meneghetti\orcid{0000-0003-1225-7084}\inst{\ref{aff10},\ref{aff16}}
\and E.~Merlin\orcid{0000-0001-6870-8900}\inst{\ref{aff23}}
\and G.~Meylan\orcid{0000-0001-6503-0209}\inst{\ref{aff71}}
\and P.~Monaco\orcid{0000-0003-2083-7564}\inst{\ref{aff72},\ref{aff12},\ref{aff13},\ref{aff11}}
\and A.~Mora\orcid{0000-0002-1922-8529}\inst{\ref{aff73}}
\and M.~Moresco\orcid{0000-0002-7616-7136}\inst{\ref{aff67},\ref{aff10}}
\and C.~Moretti\orcid{0000-0003-3314-8936}\inst{\ref{aff12},\ref{aff11},\ref{aff13}}
\and L.~Moscardini\orcid{0000-0002-3473-6716}\inst{\ref{aff67},\ref{aff10},\ref{aff16}}
\and C.~Neissner\orcid{0000-0001-8524-4968}\inst{\ref{aff74},\ref{aff26}}
\and S.-M.~Niemi\orcid{0009-0005-0247-0086}\inst{\ref{aff75}}
\and J.~W.~Nightingale\orcid{0000-0002-8987-7401}\inst{\ref{aff76}}
\and C.~Padilla\orcid{0000-0001-7951-0166}\inst{\ref{aff74}}
\and S.~Paltani\orcid{0000-0002-8108-9179}\inst{\ref{aff39}}
\and F.~Pasian\orcid{0000-0002-4869-3227}\inst{\ref{aff12}}
\and K.~Pedersen\inst{\ref{aff77}}
\and V.~Pettorino\orcid{0000-0002-4203-9320}\inst{\ref{aff75}}
\and A.~Pezzotta\orcid{0000-0003-0726-2268}\inst{\ref{aff9}}
\and S.~Pires\orcid{0000-0002-0249-2104}\inst{\ref{aff78}}
\and G.~Polenta\orcid{0000-0003-4067-9196}\inst{\ref{aff79}}
\and M.~Poncet\inst{\ref{aff80}}
\and L.~A.~Popa\inst{\ref{aff81}}
\and L.~Pozzetti\orcid{0000-0001-7085-0412}\inst{\ref{aff10}}
\and F.~Raison\orcid{0000-0002-7819-6918}\inst{\ref{aff49}}
\and R.~Rebolo\orcid{0000-0003-3767-7085}\inst{\ref{aff30},\ref{aff82},\ref{aff83}}
\and A.~Renzi\orcid{0000-0001-9856-1970}\inst{\ref{aff84},\ref{aff41},\ref{aff10}}
\and J.~Rhodes\orcid{0000-0002-4485-8549}\inst{\ref{aff53}}
\and G.~Riccio\inst{\ref{aff20}}
\and E.~Romelli\orcid{0000-0003-3069-9222}\inst{\ref{aff12}}
\and M.~Roncarelli\orcid{0000-0001-9587-7822}\inst{\ref{aff10}}
\and R.~Saglia\orcid{0000-0003-0378-7032}\inst{\ref{aff51},\ref{aff49}}
\and Z.~Sakr\orcid{0000-0002-4823-3757}\inst{\ref{aff85},\ref{aff86},\ref{aff87}}
\and D.~Sapone\orcid{0000-0001-7089-4503}\inst{\ref{aff88}}
\and B.~Sartoris\orcid{0000-0003-1337-5269}\inst{\ref{aff51},\ref{aff12}}
\and P.~Schneider\orcid{0000-0001-8561-2679}\inst{\ref{aff66}}
\and T.~Schrabback\orcid{0000-0002-6987-7834}\inst{\ref{aff89}}
\and M.~Scodeggio\inst{\ref{aff22}}
\and A.~Secroun\orcid{0000-0003-0505-3710}\inst{\ref{aff42}}
\and E.~Sihvola\orcid{0000-0003-1804-7715}\inst{\ref{aff90}}
\and P.~Simon\inst{\ref{aff66}}
\and C.~Sirignano\orcid{0000-0002-0995-7146}\inst{\ref{aff84},\ref{aff41}}
\and G.~Sirri\orcid{0000-0003-2626-2853}\inst{\ref{aff16}}
\and L.~Stanco\orcid{0000-0002-9706-5104}\inst{\ref{aff41}}
\and P.~Tallada-Cresp\'{i}\orcid{0000-0002-1336-8328}\inst{\ref{aff25},\ref{aff26}}
\and A.~N.~Taylor\inst{\ref{aff31}}
\and H.~I.~Teplitz\orcid{0000-0002-7064-5424}\inst{\ref{aff91}}
\and I.~Tereno\orcid{0000-0002-4537-6218}\inst{\ref{aff92},\ref{aff93}}
\and N.~Tessore\orcid{0000-0002-9696-7931}\inst{\ref{aff38}}
\and S.~Toft\orcid{0000-0003-3631-7176}\inst{\ref{aff94},\ref{aff95}}
\and R.~Toledo-Moreo\orcid{0000-0002-2997-4859}\inst{\ref{aff96},\ref{aff97}}
\and F.~Torradeflot\orcid{0000-0003-1160-1517}\inst{\ref{aff26},\ref{aff25}}
\and I.~Tutusaus\orcid{0000-0002-3199-0399}\inst{\ref{aff46},\ref{aff45},\ref{aff86}}
\and J.~Valiviita\orcid{0000-0001-6225-3693}\inst{\ref{aff61},\ref{aff62}}
\and T.~Vassallo\orcid{0000-0001-6512-6358}\inst{\ref{aff12},\ref{aff48}}
\and A.~Veropalumbo\orcid{0000-0003-2387-1194}\inst{\ref{aff9},\ref{aff18},\ref{aff17}}
\and Y.~Wang\orcid{0000-0002-4749-2984}\inst{\ref{aff98}}
\and J.~Weller\orcid{0000-0002-8282-2010}\inst{\ref{aff51},\ref{aff49}}
\and G.~Zamorani\orcid{0000-0002-2318-301X}\inst{\ref{aff10}}
\and F.~M.~Zerbi\orcid{0000-0002-9996-973X}\inst{\ref{aff9}}
\and E.~Zucca\orcid{0000-0002-5845-8132}\inst{\ref{aff10}}
\and S.~Borgani\orcid{0000-0001-6151-6439}\inst{\ref{aff72},\ref{aff11},\ref{aff12},\ref{aff13},\ref{aff99}}
\and A.~Loureiro\orcid{0000-0002-4371-0876}\inst{\ref{aff100},\ref{aff101}}
\and M.~Sereno\orcid{0000-0003-0302-0325}\inst{\ref{aff10},\ref{aff16}}
\and M.~Viel\orcid{0000-0002-2642-5707}\inst{\ref{aff11},\ref{aff12},\ref{aff14},\ref{aff13},\ref{aff99}}}
										   
\institute{Dipartimento di Fisica, Universit\`a degli Studi di Torino, Via P. Giuria 1, 10125 Torino, Italy\label{aff1}
\and
INFN-Sezione di Torino, Via P. Giuria 1, 10125 Torino, Italy\label{aff2}
\and
INAF-Osservatorio Astrofisico di Torino, Via Osservatorio 20, 10025 Pino Torinese (TO), Italy\label{aff3}
\and
Dipartimento di Fisica, Universit\`a di Roma Tor Vergata, Via della Ricerca Scientifica 1, Roma, Italy\label{aff4}
\and
INFN, Sezione di Roma 2, Via della Ricerca Scientifica 1, Roma, Italy\label{aff5}
\and
Universit\'e Paris-Saclay, CNRS, Institut d'astrophysique spatiale, 91405, Orsay, France\label{aff6}
\and
Universit\'e de Strasbourg, CNRS, Observatoire astronomique de Strasbourg, UMR 7550, 67000 Strasbourg, France\label{aff7}
\and
ESAC/ESA, Camino Bajo del Castillo, s/n., Urb. Villafranca del Castillo, 28692 Villanueva de la Ca\~nada, Madrid, Spain\label{aff8}
\and
INAF-Osservatorio Astronomico di Brera, Via Brera 28, 20122 Milano, Italy\label{aff9}
\and
INAF-Osservatorio di Astrofisica e Scienza dello Spazio di Bologna, Via Piero Gobetti 93/3, 40129 Bologna, Italy\label{aff10}
\and
IFPU, Institute for Fundamental Physics of the Universe, via Beirut 2, 34151 Trieste, Italy\label{aff11}
\and
INAF-Osservatorio Astronomico di Trieste, Via G. B. Tiepolo 11, 34143 Trieste, Italy\label{aff12}
\and
INFN, Sezione di Trieste, Via Valerio 2, 34127 Trieste TS, Italy\label{aff13}
\and
SISSA, International School for Advanced Studies, Via Bonomea 265, 34136 Trieste TS, Italy\label{aff14}
\and
Dipartimento di Fisica e Astronomia, Universit\`a di Bologna, Via Gobetti 93/2, 40129 Bologna, Italy\label{aff15}
\and
INFN-Sezione di Bologna, Viale Berti Pichat 6/2, 40127 Bologna, Italy\label{aff16}
\and
Dipartimento di Fisica, Universit\`a di Genova, Via Dodecaneso 33, 16146, Genova, Italy\label{aff17}
\and
INFN-Sezione di Genova, Via Dodecaneso 33, 16146, Genova, Italy\label{aff18}
\and
Department of Physics "E. Pancini", University Federico II, Via Cinthia 6, 80126, Napoli, Italy\label{aff19}
\and
INAF-Osservatorio Astronomico di Capodimonte, Via Moiariello 16, 80131 Napoli, Italy\label{aff20}
\and
Leiden Observatory, Leiden University, Einsteinweg 55, 2333 CC Leiden, The Netherlands\label{aff21}
\and
INAF-IASF Milano, Via Alfonso Corti 12, 20133 Milano, Italy\label{aff22}
\and
INAF-Osservatorio Astronomico di Roma, Via Frascati 33, 00078 Monteporzio Catone, Italy\label{aff23}
\and
INFN-Sezione di Roma, Piazzale Aldo Moro, 2 - c/o Dipartimento di Fisica, Edificio G. Marconi, 00185 Roma, Italy\label{aff24}
\and
Centro de Investigaciones Energ\'eticas, Medioambientales y Tecnol\'ogicas (CIEMAT), Avenida Complutense 40, 28040 Madrid, Spain\label{aff25}
\and
Port d'Informaci\'{o} Cient\'{i}fica, Campus UAB, C. Albareda s/n, 08193 Bellaterra (Barcelona), Spain\label{aff26}
\and
INFN section of Naples, Via Cinthia 6, 80126, Napoli, Italy\label{aff27}
\and
Institute for Astronomy, University of Hawaii, 2680 Woodlawn Drive, Honolulu, HI 96822, USA\label{aff28}
\and
Dipartimento di Fisica e Astronomia "Augusto Righi" - Alma Mater Studiorum Universit\`a di Bologna, Viale Berti Pichat 6/2, 40127 Bologna, Italy\label{aff29}
\and
Instituto de Astrof\'{\i}sica de Canarias, E-38205 La Laguna, Tenerife, Spain\label{aff30}
\and
Institute for Astronomy, University of Edinburgh, Royal Observatory, Blackford Hill, Edinburgh EH9 3HJ, UK\label{aff31}
\and
European Space Agency/ESRIN, Largo Galileo Galilei 1, 00044 Frascati, Roma, Italy\label{aff32}
\and
Universit\'e Claude Bernard Lyon 1, CNRS/IN2P3, IP2I Lyon, UMR 5822, Villeurbanne, F-69100, France\label{aff33}
\and
Institut de Ci\`{e}ncies del Cosmos (ICCUB), Universitat de Barcelona (IEEC-UB), Mart\'{i} i Franqu\`{e}s 1, 08028 Barcelona, Spain\label{aff34}
\and
Instituci\'o Catalana de Recerca i Estudis Avan\c{c}ats (ICREA), Passeig de Llu\'{\i}s Companys 23, 08010 Barcelona, Spain\label{aff35}
\and
Institut de Ciencies de l'Espai (IEEC-CSIC), Campus UAB, Carrer de Can Magrans, s/n Cerdanyola del Vall\'es, 08193 Barcelona, Spain\label{aff36}
\and
UCB Lyon 1, CNRS/IN2P3, IUF, IP2I Lyon, 4 rue Enrico Fermi, 69622 Villeurbanne, France\label{aff37}
\and
Mullard Space Science Laboratory, University College London, Holmbury St Mary, Dorking, Surrey RH5 6NT, UK\label{aff38}
\and
Department of Astronomy, University of Geneva, ch. d'Ecogia 16, 1290 Versoix, Switzerland\label{aff39}
\and
Aix-Marseille Universit\'e, CNRS, CNES, LAM, Marseille, France\label{aff40}
\and
INFN-Padova, Via Marzolo 8, 35131 Padova, Italy\label{aff41}
\and
Aix-Marseille Universit\'e, CNRS/IN2P3, CPPM, Marseille, France\label{aff42}
\and
INAF-Istituto di Astrofisica e Planetologia Spaziali, via del Fosso del Cavaliere, 100, 00100 Roma, Italy\label{aff43}
\and
INFN-Bologna, Via Irnerio 46, 40126 Bologna, Italy\label{aff44}
\and
Institut d'Estudis Espacials de Catalunya (IEEC),  Edifici RDIT, Campus UPC, 08860 Castelldefels, Barcelona, Spain\label{aff45}
\and
Institute of Space Sciences (ICE, CSIC), Campus UAB, Carrer de Can Magrans, s/n, 08193 Barcelona, Spain\label{aff46}
\and
School of Physics, HH Wills Physics Laboratory, University of Bristol, Tyndall Avenue, Bristol, BS8 1TL, UK\label{aff47}
\and
University Observatory, LMU Faculty of Physics, Scheinerstr.~1, 81679 Munich, Germany\label{aff48}
\and
Max Planck Institute for Extraterrestrial Physics, Giessenbachstr. 1, 85748 Garching, Germany\label{aff49}
\and
INAF-Osservatorio Astronomico di Padova, Via dell'Osservatorio 5, 35122 Padova, Italy\label{aff50}
\and
Universit\"ats-Sternwarte M\"unchen, Fakult\"at f\"ur Physik, Ludwig-Maximilians-Universit\"at M\"unchen, Scheinerstr.~1, 81679 M\"unchen, Germany\label{aff51}
\and
Institute of Theoretical Astrophysics, University of Oslo, P.O. Box 1029 Blindern, 0315 Oslo, Norway\label{aff52}
\and
Jet Propulsion Laboratory, California Institute of Technology, 4800 Oak Grove Drive, Pasadena, CA, 91109, USA\label{aff53}
\and
Felix Hormuth Engineering, Goethestr. 17, 69181 Leimen, Germany\label{aff54}
\and
Technical University of Denmark, Elektrovej 327, 2800 Kgs. Lyngby, Denmark\label{aff55}
\and
Cosmic Dawn Center (DAWN), Denmark\label{aff56}
\and
Max-Planck-Institut f\"ur Astronomie, K\"onigstuhl 17, 69117 Heidelberg, Germany\label{aff57}
\and
NASA Goddard Space Flight Center, Greenbelt, MD 20771, USA\label{aff58}
\and
Department of Physics and Astronomy, University College London, Gower Street, London WC1E 6BT, UK\label{aff59}
\and
Universit\'e de Gen\`eve, D\'epartement de Physique Th\'eorique and Centre for Astroparticle Physics, 24 quai Ernest-Ansermet, CH-1211 Gen\`eve 4, Switzerland\label{aff60}
\and
Department of Physics, P.O. Box 64, University of Helsinki, 00014 Helsinki, Finland\label{aff61}
\and
Helsinki Institute of Physics, Gustaf H{\"a}llstr{\"o}min katu 2, University of Helsinki, 00014 Helsinki, Finland\label{aff62}
\and
Laboratoire d'etude de l'Univers et des phenomenes eXtremes, Observatoire de Paris, Universit\'e PSL, Sorbonne Universit\'e, CNRS, 92190 Meudon, France\label{aff63}
\and
SKAO, Jodrell Bank, Lower Withington, Macclesfield SK11 9FT, UK\label{aff64}
\and
Centre de Calcul de l'IN2P3/CNRS, 21 avenue Pierre de Coubertin 69627 Villeurbanne Cedex, France\label{aff65}
\and
Universit\"at Bonn, Argelander-Institut f\"ur Astronomie, Auf dem H\"ugel 71, 53121 Bonn, Germany\label{aff66}
\and
Dipartimento di Fisica e Astronomia "Augusto Righi" - Alma Mater Studiorum Universit\`a di Bologna, via Piero Gobetti 93/2, 40129 Bologna, Italy\label{aff67}
\and
Department of Physics, Institute for Computational Cosmology, Durham University, South Road, Durham, DH1 3LE, UK\label{aff68}
\and
Universit\'e Paris Cit\'e, CNRS, Astroparticule et Cosmologie, 75013 Paris, France\label{aff69}
\and
CNRS-UCB International Research Laboratory, Centre Pierre Bin\'etruy, IRL2007, CPB-IN2P3, Berkeley, USA\label{aff70}
\and
Institute of Physics, Laboratory of Astrophysics, Ecole Polytechnique F\'ed\'erale de Lausanne (EPFL), Observatoire de Sauverny, 1290 Versoix, Switzerland\label{aff71}
\and
Dipartimento di Fisica - Sezione di Astronomia, Universit\`a di Trieste, Via Tiepolo 11, 34131 Trieste, Italy\label{aff72}
\and
Telespazio UK S.L. for European Space Agency (ESA), Camino bajo del Castillo, s/n, Urbanizacion Villafranca del Castillo, Villanueva de la Ca\~nada, 28692 Madrid, Spain\label{aff73}
\and
Institut de F\'{i}sica d'Altes Energies (IFAE), The Barcelona Institute of Science and Technology, Campus UAB, 08193 Bellaterra (Barcelona), Spain\label{aff74}
\and
European Space Agency/ESTEC, Keplerlaan 1, 2201 AZ Noordwijk, The Netherlands\label{aff75}
\and
School of Mathematics, Statistics and Physics, Newcastle University, Herschel Building, Newcastle-upon-Tyne, NE1 7RU, UK\label{aff76}
\and
DARK, Niels Bohr Institute, University of Copenhagen, Jagtvej 155, 2200 Copenhagen, Denmark\label{aff77}
\and
Universit\'e Paris-Saclay, Universit\'e Paris Cit\'e, CEA, CNRS, AIM, 91191, Gif-sur-Yvette, France\label{aff78}
\and
Space Science Data Center, Italian Space Agency, via del Politecnico snc, 00133 Roma, Italy\label{aff79}
\and
Centre National d'Etudes Spatiales -- Centre spatial de Toulouse, 18 avenue Edouard Belin, 31401 Toulouse Cedex 9, France\label{aff80}
\and
Institute of Space Science, Str. Atomistilor, nr. 409 M\u{a}gurele, Ilfov, 077125, Romania\label{aff81}
\and
Consejo Superior de Investigaciones Cientificas, Calle Serrano 117, 28006 Madrid, Spain\label{aff82}
\and
Universidad de La Laguna, Dpto. Astrof\'\i sica, E-38206 La Laguna, Tenerife, Spain\label{aff83}
\and
Dipartimento di Fisica e Astronomia "G. Galilei", Universit\`a di Padova, Via Marzolo 8, 35131 Padova, Italy\label{aff84}
\and
Instituto de F\'isica Te\'orica UAM-CSIC, Campus de Cantoblanco, 28049 Madrid, Spain\label{aff85}
\and
Institut de Recherche en Astrophysique et Plan\'etologie (IRAP), Universit\'e de Toulouse, CNRS, UPS, CNES, 14 Av. Edouard Belin, 31400 Toulouse, France\label{aff86}
\and
Universit\'e St Joseph; Faculty of Sciences, Beirut, Lebanon\label{aff87}
\and
Departamento de F\'isica, FCFM, Universidad de Chile, Blanco Encalada 2008, Santiago, Chile\label{aff88}
\and
Universit\"at Innsbruck, Institut f\"ur Astro- und Teilchenphysik, Technikerstr. 25/8, 6020 Innsbruck, Austria\label{aff89}
\and
Department of Physics and Helsinki Institute of Physics, Gustaf H\"allstr\"omin katu 2, University of Helsinki, 00014 Helsinki, Finland\label{aff90}
\and
Infrared Processing and Analysis Center, California Institute of Technology, Pasadena, CA 91125, USA\label{aff91}
\and
Departamento de F\'isica, Faculdade de Ci\^encias, Universidade de Lisboa, Edif\'icio C8, Campo Grande, PT1749-016 Lisboa, Portugal\label{aff92}
\and
Instituto de Astrof\'isica e Ci\^encias do Espa\c{c}o, Faculdade de Ci\^encias, Universidade de Lisboa, Tapada da Ajuda, 1349-018 Lisboa, Portugal\label{aff93}
\and
Cosmic Dawn Center (DAWN)\label{aff94}
\and
Niels Bohr Institute, University of Copenhagen, Jagtvej 128, 2200 Copenhagen, Denmark\label{aff95}
\and
Universidad Polit\'ecnica de Cartagena, Departamento de Electr\'onica y Tecnolog\'ia de Computadoras,  Plaza del Hospital 1, 30202 Cartagena, Spain\label{aff96}
\and
European University of Technology EUt+, European Union\label{aff97}
\and
Caltech/IPAC, 1200 E. California Blvd., Pasadena, CA 91125, USA\label{aff98}
\and
ICSC - Centro Nazionale di Ricerca in High Performance Computing, Big Data e Quantum Computing, Via Magnanelli 2, Bologna, Italy\label{aff99}
\and
Oskar Klein Centre for Cosmoparticle Physics, Department of Physics, Stockholm University, Stockholm, SE-106 91, Sweden\label{aff100}
\and
Astrophysics Group, Blackett Laboratory, Imperial College London, London SW7 2AZ, UK\label{aff101}}    

%
%
\abstract{
The cosmic infrared background, sourced by dust heated by star-forming activity, encodes the integrated history of the star-formation distribution over a large swathe of cosmic time. Decomposing the contribution from galaxies at different redshifts to the observed CIB maps provides a powerful tool for probing the large-scale star-formation history across cosmic times. To this end, we perform a forecast tomographic cross-correlation analysis using simulated photometric galaxy clustering and weak lensing data based on the \Euclid mission survey specifications, combined with \Planck CIB observations. The analysis adopts templates constructed from the halo occupation distribution model to measure the bias-weighted star-formation-rate density, \brsfrd, as a function of redshift. We present forecasts for the \brsfrd\ constraints based on these simulations and quantify the expected constraining power using Fisher information matrix methods. Compared with results obtained using the same dataset as in previous works, we find that \Euclid will tighten the constraints on \brsfrd\ across all redshift bins by a factor of about $2$ and provide more measurements over a deeper redshift range. This leads to improved constraints on all HOD parameters, typically reducing the marginalized uncertainties by a factor of about $2.5$, while also breaking parameter degeneracies.
}
%
%

\titlerunning{Star-formation history with \Euclid-CIB cross-correlations}
\authorrunning{Han et al.}
   
\maketitle
%
%
%
%
   
\section{\label{sc:Intro}Introduction}
The cosmic infrared background (CIB) consists of the integrated emission from unresolved dusty star-forming galaxies. The dust grains in the galaxies absorb the ultraviolet radiation from young stars and re-emit it in the infrared. The majority of the CIB emission arises from the peak of the star-formation epoch, occurring at redshifts $z \approx 1$--$2$, commonly known as cosmic noon, and is predominantly associated with galaxies residing in dark matter haloes with masses ranging from $10^{11}$ to $10^{13}\,M_\odot$ \citep{Bethermin:2012ki,Schmidt:2014jja}. Anisotropies in the CIB encode information about the spatial distribution of galaxies and their underlying dark matter content.

The study of the CIB has a rich history, dating back to its first detection in the diffuse infrared sky background in the 1990s by the DIRBE and FIRAS instruments on board the \textit{COBE} satellite. In the following decades, the anisotropies of the CIB were detected, enabling measurements of the CIB auto-power spectrum and constraining the effective large-scale bias of the CIB sources relative to the dark matter distribution. Significant progress in understanding the CIB was achieved through the \Planck and \textit{Herschel} satellite missions, with relevant contributions from ground-based experiments (\citealt{2010Hall, ade2011planck, PlanckCIB2014, PlanckCIBremazeille, 2013Viero, 2019Viero}). 

Despite being a nuisance foreground contaminant in cosmic microwave background (CMB) observations, the CIB serves as a powerful tool for both cosmological and astrophysical studies. In cosmology, it provides a means to probe the large-scale structure \citep{puget1996tentative,gispert2000implications,lagache2005dusty,dole2006cosmic} and can be used to (partially) remove the effects of gravitational lensing on the CMB \citep{ade2014planck,sherwin2015delensing}. In astrophysics, the CIB is employed as a highly sensitive tracer of the star-formation rate (SFR), since its emission originates from dust heated by newly formed stars \citep{dole2006cosmic}. This makes it particularly valuable for studying star formation across time, especially given the observational challenges of collecting large samples of star-forming galaxies at high redshifts. Understanding star formation across time and galaxy type is fundamental for unravelling the processes of galaxy formation and evolution \citep[see][for a comprehensive study of star formation and galaxy evolution]{Madau2014,vogelsberger2020cosmological}.

Unfortunately, it is extremely difficult to disentangle the contribution from sources at different redshifts to the CIB using observations of the CIB signal alone because of its projected nature. This issue can be at least partly bypassed by cross-correlating the CIB with the distribution of galaxies at different redshifts -- for example, the cosmological probe of galaxy clustering -- since the sources from which the CIB emission originated trace the same large-scale structure as the galaxy sample. \citet{2022A&A...665A..52Y} and \citet{Jego:2022eqo} showed that large-scale galaxy--CIB cross-correlations allow a tomographic determination of the bias-weighted star-formation rate density (SFRD), \brsfrd, without assuming any explicit SFR model, as is directly inferred from the angular power spectra.  Furthermore, \citet{Jego:2022fkl} investigated the cross-correlation between CIB and tomographic weak lensing measurements. Since the cosmic shear signal from galaxies at different redshifts directly traces matter inhomogeneities, this cross-correlation is sensitive to the relation between the SFR and matter densities at different times.

In this work, we extend the previous analyses to the \Euclid mission, a medium-class project of the European Space Agency (ESA) focused on exploring dark matter and dark energy, as well as galaxy formation and evolution \citep{Laureijs11,EuclidSkyOverview}. \Euclid will survey about one-third of the sky, conducting one of the largest galaxy surveys ever undertaken. To study the Universe's expansion history, it will use two main cosmological probes: weak gravitational lensing, which measures the distortions of galaxy shapes caused by matter; and galaxy clustering, which traces the large-scale distribution of galaxies as biased tracers of the underlying matter density field.

Here, we set out to assess how the cross-correlation of the CIB with both tomographic cosmic shear measurements and tomographic galaxy clustering data from \Euclid can reconstruct the star-formation history. First, we apply an HOD method to derive theoretical predictions for the galaxy clustering auto-power spectrum using the \Euclid tomographic photometric redshift bins, and for the cross-power spectrum between CIB and galaxy clustering, using CIB observations from \Planck \citep{Lenz:2019ugy, PlanckCIBremazeille}. 
We perform an information matrix analysis to assess the constraining power of this method. Then, we perform a similar analysis of the cross-correlation between CIB and weak lensing using the galaxy number density distribution across the \Euclid mission's photometric bins.

In this paper, we assume a flat concordance cosmological model with the following fiducial values for the model parameters: $\Omega_{\rm m}=0.32$, $\Omega_{\rm b}=0.05$, $h=0.67$, $n_{\rm s}=0.96$, $\sigma_8=0.816$, $\sum m_{\nu}=0.06\,\mathrm{eV}$, $\tau=0.058$ \citep{Planck:2018vyg}.

This paper is organized as follows. In \cref{sc:Meth}, we introduce the theoretical framework used to model projected cosmological observables within the halo model, and describe the formalism for galaxy clustering, cosmic shear, and CIB anisotropies, including the treatment of intrinsic alignments. In \cref{tomo}, we present the tomographic template-fitting methodology used to extract \brsfrd\ from CIB cross-correlations with \Euclid galaxy clustering and weak lensing data and discuss the implementation of the Bernardeau--Nishimichi--Taruya (BNT) nulling technique for shear \citep*{bernardeau2014cosmic}. Our forecast results and their interpretation are presented in \cref{res}, where we assess the expected constraints on the star-formation history and halo-model parameters, and compare them with current Dark Energy Spectroscopic Instrument (DESI) Legacy Imaging Surveys (hereafter DELS) CIB analyses. Finally, we summarise our conclusions and outline future prospects in \cref{sc:conc}.

\section{\label{sc:Meth}Modelling of the observables}
Our method follows the formalism described in various papers in the literature \citep[see e.g.][]{moster2019emerge,garcia2021growth,Jego:2022fkl,Jego:2022eqo}, which we shall briefly review here.


\subsection{\label{probe}Halo model for harmonic-space power spectra}
Let $u$ represent a cosmological field projected onto the celestial sphere, assumed to result from the radial projection of a 3D field, $U$. If \(\vec r=r\,\hat{\vec r}\) is a 3D comoving position vector, with magnitude \(r=|\vec r|\) and direction \(\hat{\vec r}\), we have
\begin{equation}
    u(\hat{\vec r})=\int^{z_{\rm hi}}_{z_{\rm lo}}\frac{c\,\de z}{H(z)}\, q_u(z)\,U(\vec r, z)\,,
    \label{eq:un}
\end{equation}
where $q_u$ denotes the radial kernel associated with $u$, and $z$ is the redshift corresponding to the comoving distance $r$, such that $\de r=c\,\de z / H(z)$, and $H(z)$ is the Hubble expansion rate as a function of the redshift. The line-of-sight integration limits $[z_{\rm lo}, z_{\rm hi}]$ are set separately for each tracer. For galaxy clustering and cosmic shear, we adopt $[0, 3]$, following the redshift distribution of the photometric galaxy sample. For the CIB, the range is extended to $[0, 10]$, as dictated by the spectral energy distributions of \citet{maniyar2021simple}. For the cross-correlation between two tracers, we integrate over the overlap of their respective kernels, which yields $[0, 3]$. The angular power spectrum of two such projected quantities, $u$ and $v$, denoted as $C^{uv}(\ell)$ and defined as the covariance of their harmonic-expansion coefficients, is related to the power spectrum of their corresponding 3D source fields, $P_{UV}(k,z)$, through a similar radial projection, that is,
\begin{equation}
    C^{uv}(\ell)=\int^{z_{\rm hi}}_{z_{\rm lo}}\frac{c\,\de z}{H(z)}\,\frac{q_u(z)\,q_v(z)}{r^2(z)}\,P_{U V}\left[\frac{\ell+1/2}{r(z)}, z\right]\,.
    \label{eq:cell}
\end{equation}

Here, we used the so-called Limber approximation setting $k=(\ell+1/2)/r(z)$, which is valid when the radial kernels are significantly broader than the typical correlation length of the fields \citep{kaiser1998weak,loverde2008extended}. This condition holds, in general, for $\ell\gg1$, and integrated cosmological observables such as weak lensing tend to be very well described by the Limber approximation. On the other hand, galaxy clustering, with its narrower kernels, is known to deviate significantly from the naive Limber prediction even at multipoles of a few tens, which is usually considered safe for these kinds of analyses \citep{hang2021galaxy,garcia2021growth}. For this reason, here we adopt a very conservative cut at $\ell_{\rm min}=100$ \citep{2019MNRAS.489.3385T,Tanidis-TBD}. As a sanity check, we verified that lowering this cut to $\ell_{\rm min}=60$ leads to negligible changes in our results, confirming that our conclusions are not sensitive to the precise choice of $\ell_{\rm min}$ within this range.

For the source fields, we employ the halo model \citep{seljak2000analytic, peacock2000halo,cooray2002halo}, which is a versatile formalism for modeling 3D correlations of cosmological quantities. In this framework, the (dark) matter power spectrum is expressed as the sum of two components: a two-halo term, describing the clustering of dark matter haloes on cosmological scales; and a one-halo term, accounting for correlations within the same dark matter halo. When describing two different tracers of the dark matter distribution, $U$ and $V$, we can write $P_{UV}(k)=P^{\rm1h}_{UV}(k)+P^{\rm2h}_{UV}(k)$,\footnote{We omit the redshift dependence to avoid clutter.} with
\begin{align}
    P^{\rm1h}_{UV}(k) &= \int^{M_{\rm hi}}_{M_{\rm lo}}\de M \, n_{\rm h}(M) \, \langle U(k, M)\,V(k, M) \rangle\,, \label{eq:1halo} \\
    P^{\rm2h}_{UV}(k) &= \langle bU(k) \rangle\,\langle bV(k) \rangle\,P_\text{lin}(k)\,. \label{eq:2halo}
\end{align}
Here, $U(k, M)$ denotes the Fourier-space profile of the quantity $U(\vec r)$ in a halo of mass $M$, $n_{\rm h}(M)$ represents the halo mass function, describing the comoving number density of haloes per unit mass interval, and $P_\text{lin}(k)$ is the linear matter power spectrum. All halo-mass integral in our analysis are evaluated over upper limit $M_{\rm hi}=10^{16}M_\odot$ and lower limit $M_{\rm lo}=10^8M_\odot$, we verified that our results are unchanged if these limits are widened. Then, quantities such as $\langle bU(k) \rangle$ represent the bias-weighted values of the function $U$, and are defined as
\begin{equation}
    \langle bU(k) \rangle \coloneqq \int^{M_{\rm hi}}_{M_{\rm lo}}\de M \, n_{\rm h}(M) \, b_{\rm h}(M) \, \langle U(k, M) \rangle\,, \label{eq:bU}
\end{equation}
where $b_{\rm h}(M)$ is the bias of a halo of mass $M$.

On sufficiently large scales, both $P^{\rm1h}_{UV}$ and $\langle bU \rangle$ tend to a constant. Thus, on those scales, the power spectrum can be approximated as
\begin{equation}
    P_{UV}(k) \simeq \langle bU \rangle\,\langle bV \rangle P_\text{lin}(k) + N_{UV}\,, \label{eq:large_scale}
\end{equation}
where $N_{UV}$ denotes the scale-independent contribution from the one-halo term. However, over the range of scales considered in this work, $k<0.3\,h\,{\rm Mpc}^{-1}$, the constant one-halo contribution is negligible compared to the clustering term, such that
$N_{UV}\ll \langle bU\rangle\langle bV\rangle P_{\rm lin}(k)$. Accordingly, only the two-halo contribution is used to model the angular power spectra. The shot-noise term is included solely in the covariance matrix calculation and is shown separately in Fig.~\ref{fig:ps} for comparison. As discussed in \citet{Jego:2022eqo}, extending the analysis to smaller scales would require a more detailed treatment of nonlinear and one-halo contributions. We therefore focus on large scales, enabling a robust and largely model-independent measurement of the bias-weighted quantities $\langle bU\rangle$, which can subsequently be used to constrain the cosmic star-formation history.

\subsection{\label{GC}Cosmological probes}
We start by reviewing the formalism for \Euclid's cosmological probes, i.e., galaxy clustering and cosmic shear. The projected galaxy overdensity field \(\delta_{\rm g}(\hat{\vec r})\) is related to its 3D counterpart \(\Delta(\vec r)=b_{\rm g}\,\delta(\vec r)\) via \cref{eq:un}, with $b_{\rm g}$ the linear galaxy bias and $\delta$ the matter density contrast, by the kernel
\begin{equation}
    q_{\rm g}(z)=\frac{H(z)}{c}\,p(z)\,.
    \label{eq:delta}
\end{equation}
Here, $p(z)=n(z)/\bar n$ represents the probability of finding a galaxy between redshift $z$ and $z+\de z$, with $n(z)$ the galaxy redshift distribution (namely the number of galaxies per unit redshift per steradian) and $\bar n\coloneqq\int\de z\,n(z)$ is the mean galaxy number density. The associated $\langle bU \rangle$ is then simply $b_{\rm g}$. In this study, we treat $b_{\rm g}$ as a free nuisance parameter. 

The clustering signal, in fact, contains contributions beyond the dominant one proportional to density fluctuations \citep[see e.g.,][]{kaiser1987clustering,schneider1992gravitational,yoo2009fast}, the most important of which are redshift-space distortions and lensing magnification \citep{challinor2011linear,bonvin2011galaxy}. 
The magnification contribution can be modeled by writing the observed galaxy overdensity as
$\delta_{\rm obs}=\delta_{\rm g}+(5s-2)\kappa$
\citep{Turner1984,Loverde2008}, where $s$ is the slope of the cumulative galaxy number counts and $\kappa$ is the lensing convergence. 
While magnification bias can have a non-negligible impact on galaxy number-count correlations \citep[e.g.,][]{Lepori-EP19}, its contribution to the galaxy--CIB cross-correlation considered here is expected to be subdominant. This is because the dominant contribution to the magnification term arises from structures at relatively low redshift where the overlap with the CIB kernel is limited. Using the same theoretical framework adopted for the clustering signal, we explicitly write the magnification contribution as \citep{Lepori-EP19}:
\begin{equation}
    s(z) = s_0 + s_1 z + s_2 z^2 + s_3 z^3\,,
\end{equation}
where $s_0 = 0.0842, \,s_1 = 0.0532,\, s_2 = 0.298,\,s_3 = -0.0113$.

We evaluate the impact of magnification bias and find that its contribution to the galaxy--CIB cross-correlation is generally below the $5\%$ level over the angular scales considered in this work. The contribution increases in the deepest redshift bin, reaching approximately $8\%$, due to increased lensing efficiency at higher redshift. However, this remains subdominant compared with the expected statistical uncertainties of the measurement, and we therefore neglect magnification bias in the galaxy--CIB analysis. 
For weak-lensing shear correlations, magnification bias enters as a subdominant correction to the measured signal. \citet{Deshpande20} showed that its impact on \Euclid-like shear correlation functions is expected to remain small over the angular scales relevant for cosmological analyses. We therefore neglect this contribution in the shear analysis as well.
Redshift-space distortions, on the other hand, have been shown to be important for \Euclid's photometric probes \citep{2019MNRAS.489.3385T,Tanidis-TBD}, but they mainly affect the largest scales. Given our conservative choice of $\ell_{\rm min}=100$, we can therefore also safely neglect them.

\Euclid's other main cosmological probe is the weak gravitational lensing effect of cosmic shear, which distorts the shapes of background galaxies, resulting in correlated ellipticities. This effect can be described as a spin-2 projected field $\gamma(\hat{\vec r})$ sourced by the unbiased matter density distribution $\delta(\vec r)$, with radial kernel\footnote{Note that in principle there is an $\ell$-dependent prefactor in the shear kernel, stemming from the relation between $\gamma$ and the angular Hessian of the Newtonian gravitational potential. However, this term is important only on large angular scales, and at the minimum multipole adopted in our analysis, the deviation from unity amounts to less than $0.1\,\%$. Therefore, we shall neglect it in what follows.}
\begin{equation}
    q_\gamma(z)=\frac{3}{2}\,\frac{H_0^2}{c^2}\,\Omega_{\rm m}\,(1+z)\,r(z)\,\int_z^{\infty}\de z^{\prime}\,p\left(z^{\prime}\right)\,\frac{r(z^{\prime})-r(z)}{r(z^{\prime})}\,.
    \label{eq:qgamma}
\end{equation}

\begin{figure}[htbp!]
\centering
 \includegraphics[angle=0,width=1\columnwidth]{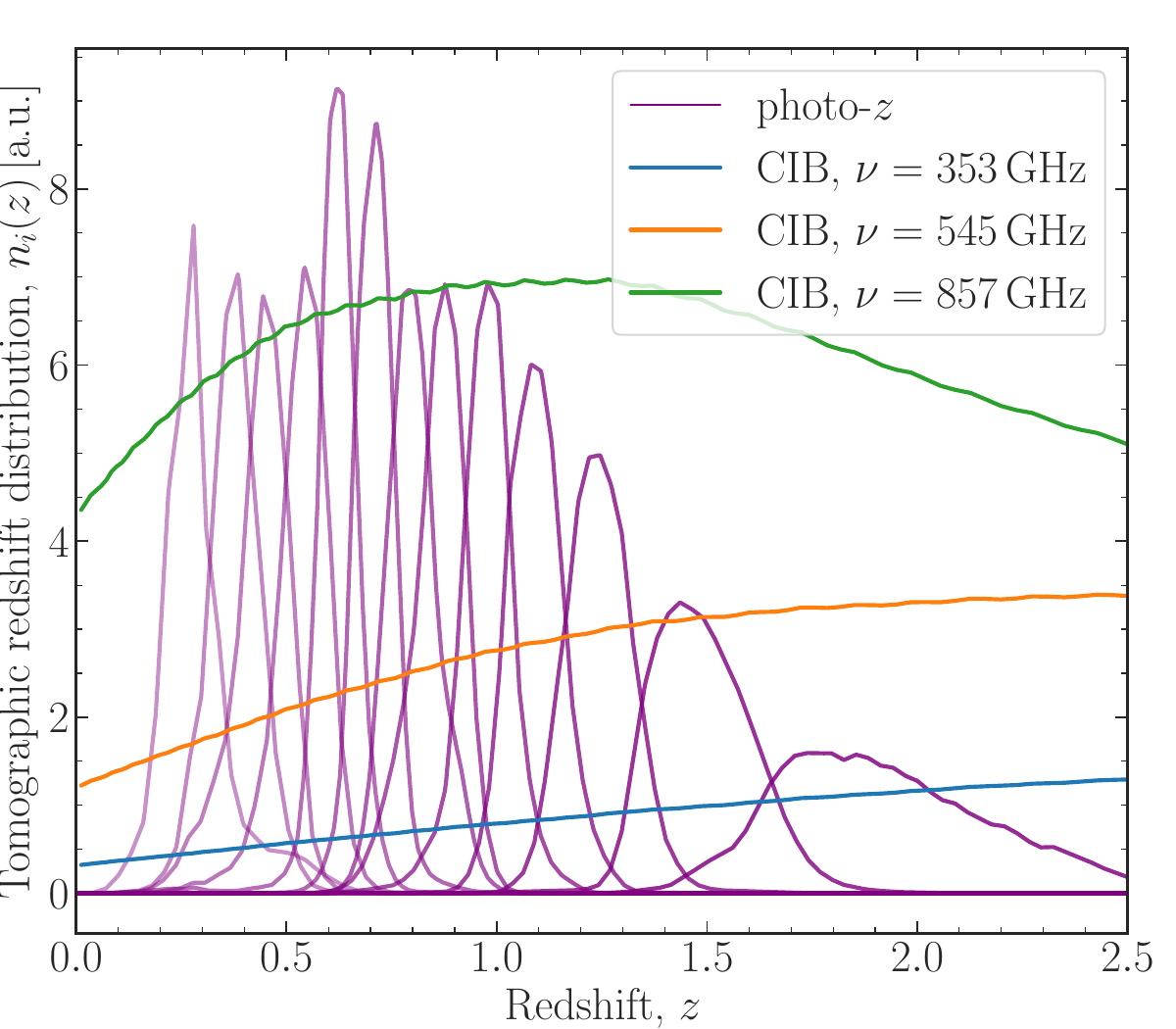}
\caption{Redshift distribution of galaxies in the \Euclid photometric sample. These are shown in purple, divided into 13 equi-populated bins. The kernels for the CIB at $353$, $545$, and $857$\,GHz are shown in blue, orange, and green, respectively.}
\label{fig:zker}
\end{figure}

Cosmic shear is estimated from the correlation in galaxy ellipticities. However, such correlations also include a contribution due to so-called intrinsic alignments (IAs), which arise from the influence of local gravitational tidal fields \citep{larsen2016intrinsic} and can introduce a significant contaminant in weak lensing measurements \citep{brown2002measurement}. Unlike cosmic shear, which traces the integrated density field of the cosmic large-scale structure, IAs are localized and correlate with the structures at the redshifts of the source galaxies. 

Following \citet{Blanchard-EP7} and \citet{paganin2024euclid}, we include the effect of IAs as a local contribution to the cosmic shear kernel, namely
\begin{equation}
q_{\rm IA}(z) 
= - \mathcal{A}_{\mathrm{IA}} \, C_{\mathrm{IA}} \, \Omega_{\rm m}\,\frac{\mathcal{F}_{\mathrm{IA}}(z)}{D(z)} \,
  p(z)\frac{H(z)}{c}\,,
\end{equation}
with $D(z)$ being the linear growth factor, while the function $\mathcal{F}_{\rm IA}(z)$ reads
\begin{equation}
    \mathcal{F}_{\mathrm{IA}}(z)=(1+z)^{\eta_{\mathrm{IA}}}\left[\langle L\rangle(z) / L_{\star}(z)\right]^{\beta_{\mathrm{IA}}}\,.
\end{equation}
Here, $\langle L \rangle(z)$ and $L_{\ast}(z)$ are the redshift-dependent mean and characteristic luminosities of the source galaxies, as computed from the luminosity function. In our analysis, we adopt the linear/nonlinear alignment model \citep[][]{hirata2004intrinsic}, following \citet{Jego:2022fkl}, so that the luminosity dependence is ignored by making $\beta_{\rm IA}=0$. 
We adopt the fiducial values $\{\mathcal{A}_{\rm IA},\,\eta_{\rm IA},\,C_{\rm IA}\}=\{1.72,\,-0.41,\,0.0134\}$ \citep{paganin2024euclid}, which specify the fiducial model used to compute the theoretical observables and their derivatives entering the Fisher matrix. The parameters $\mathcal{A}_{\rm IA}$ and $\eta_{\rm IA}$ are treated as nuisance parameters and marginalized over in the forecast, whereas $C_{\rm IA}$ is fixed throughout the analysis since it is degenerate with $\mathcal{A}_{\rm IA}$.
As a result, the kernel for the measured ellipticity field ends up being
\begin{equation}
    q_\epsilon(z)=q_\gamma(z)+q_{\rm IA}(z)\,.
\end{equation}

For the \Euclid tracers in our analysis, we adopt a tomographic binning of the \Euclid\ photometric galaxy sample into 13 redshift bins, following the configuration used in \Euclid forecast studies \citep[][]{Laureijs11,Pocino-EP12}, which is shown in \cref{fig:zker}.

\subsection{\label{CIB}CIB}
Let us now move to the main focus of this work, which is the CIB as a tracer of SFR history. The CIB intensity at a given observed frequency, $\nu$, is connected to the infrared emissivity $j_{\nu(1+z)}$ (energy emitted per unit frequency, time, and comoving volume in the emitter's rest frame), through the relation
\begin{equation}
    I_\nu(\hat{\vec r}) = \int^{z_{\rm hi}}_{z_{\rm lo}} \frac{c\,\de z}{H(z)}\,\frac{j_{\nu(1+z)}(\hat{\vec r}, z)}{4\,\pi\,(1+z)}\,.
    \label{eq:Inu}
\end{equation}
Assuming a model for the specific infrared luminosity of haloes with mass $M$, denoted as $L_\nu(M)$, the mean comoving emissivity can be expressed as
\begin{equation}
    \langle j_{\nu(1+z)} \rangle = \int^{M_{\rm hi}}_{M_{\rm lo}}\de M \, n_{\rm h}(M)\,L_{\nu(1+z)}(M)\,.
    \label{eq:jnu}
\end{equation}
The specific infrared luminosity can be related to the observed source flux density $S_{\nu}$ as
\begin{equation}
    L_{\nu(1+z)}(z)= 4\,\pi\,r^2\,(1+z)^2\,S_\nu\,.
    \label{eq:Lnu}
\end{equation}
Then, the total infrared luminosity, $L_\mathrm{IR}$, is highly correlated with the source SFR,
\begin{equation}
    \sfr = K\,L_\mathrm{IR}\,,
    \label{eq:SFR}
\end{equation}
where $K = 10^{-10}\,M_\odot \, \mathrm{yr}^{-1} \, L_\odot^{-1}$ is the calibration constant between the far-infrared luminosity and the SFR \citep{kennicutt1998star,kennicutt2012star} for a Chabrier initial mass function \citep{chabrier2003galactic}, and where the total infrared luminosity $L_\mathrm{IR}$ is integrated over the range $8$--$\qty{1000}{\micro\meter}$.

By combining \cref{eq:jnu,eq:Lnu,eq:SFR}, the CIB intensity can be expressed as
\begin{equation}
     I_\nu(\hat{\vec r}) = \frac1K\,\int^{z_{\rm hi}}_{z_{\rm lo}}\frac{c\,\de z}{H(z)}\,r^2(z)\,S_\nu^\mathrm{eff}(z)\,\rsfrd(\hat{\vec r}, z)\,.
    \label{eq:Inu_d}
\end{equation}
Here, $\rsfrd$ represents the SFRD, given by the contribution of all haloes of different masses, while $S_\nu^\mathrm{eff}(z)$ corresponds to $S_\nu(z)/L_\mathrm{IR}$, the mean flux density of sources at redshift $z$ normalized to the total luminosity. We specifically use the estimates of $S_\nu^\mathrm{eff}(z)$ for the $545\,\mathrm{GHz}$ \Planck frequency channel \citep[][hereafter M21]{maniyar2021simple}.

As in \cref{eq:un}, the CIB intensity $I_\nu(\hat{\vec r})$ can be interpreted as a projected tracer of $\rsfrd$, characterised by the radial kernel
\begin{equation}
    q_\nu(z) = \frac{r^2(z)}{K}\,S_\nu^\mathrm{eff}(z)\,.
    \label{eq:qnu}
\end{equation}
Then, in the case of CIB anisotropies, the quantity $\langle bU \rangle$ associated with it is the mean SFRD weighted by the halo bias, i.e.,
\begin{equation}
    \brsfrd\coloneqq\int^{M_{\rm hi}}_{M_{\rm lo}}\de M \, n_{\rm h}(M) \, b_{\rm h}(M) \, \sfr(M)\,.\label{eq:brsfr}
\end{equation}
Here, $\sfr(M, z)$ is the mean SFR in haloes of mass $M$ at redshift $z$. 

To describe the \sfr-mass relation, we adopt a halo-model parameterization from \citet{maniyar2018star} and \citetalias{maniyar2021simple}. This model separates the total \sfr\ of a given halo into contributions from central ($\sfr_\mathrm{c}$) and satellite galaxies ($\sfr_\mathrm{s}$), namely,
\begin{equation}
    \sfr(M, z) = \sfr_\mathrm{c}(M, z) + \sfr_\mathrm{s}(M, z)\,.\label{eq:sfr}
\end{equation}
For central galaxies, the process of gas accretion, with a fraction being converted into stars, can be modeled as
\begin{equation}
    \sfr_\mathrm{c}(M, z) = \eta(M, z) \, \mathrm{BAR}(M, z)\,,\label{eq:sfr_c}
\end{equation}
where $\mathrm{BAR}$ denotes the baryon-accretion rate and $\eta$ represents the efficiency of converting accreted baryons into stars. We model $\mathrm{BAR}$ after \citet{2010MNRAS.406.2267F} as
\begin{equation}
    \mathrm{BAR}(M, z) = \dot{M_0}\,\frac{\Omega_{\rm b}}{\Omega_{\rm m}} \, \left(\frac{M}{10^{12}\,M_\odot}\right)^{1.1}\,\frac{(1 + 1.11\,z)}{H_0/H(z)}\,,\label{eq:bar}
\end{equation}
where $\dot{M_0} = 46.1\,M_\odot \, \mathrm{yr}^{-1}$. For efficiency, we follow \citetalias{maniyar2021simple},
\begin{equation}
    \eta(M, z) = \eta_{\max} \,\exp\left[-\frac{\log_{10}({M}/{M_{\max}})}{2\,\sigma^2_M(M,z)}\right]\,,\label{eq:eta}
\end{equation}
where $M_{\max}$ represents the halo mass at which the star-formation efficiency, $\eta_{\max}$, reaches its maximum. The remaining quantities define the range of halo masses with efficient star formation, which varies as a function of both mass and redshift, in particular
\begin{equation}
   \sigma_{M}(M, z)=\sigma_{0}-\tau 
    \,\max \left(0, z_{\rm c}-z\right)\,.\label{eq:sigma_M}
\end{equation}
The model in \citetalias{maniyar2021simple}, with $z_{\rm c}$ fixed at 1.5, is defined by four free parameters: $\eta_{\max}$, $\log_{10} (M_{\max}/M_{\odot})$, $\sigma_{0}$, and $\tau$. Here, we constructed our model based on the best-fit values $\log_{10} (M_{\max}/M_{\odot}) = 12.94$, $\eta_{\max} = 0.42$, $\sigma_{0} = 1.75$, and $\tau = 1.17$.

\begin{figure}[htbp!]
\centering
\includegraphics[angle=0,width=1\columnwidth]{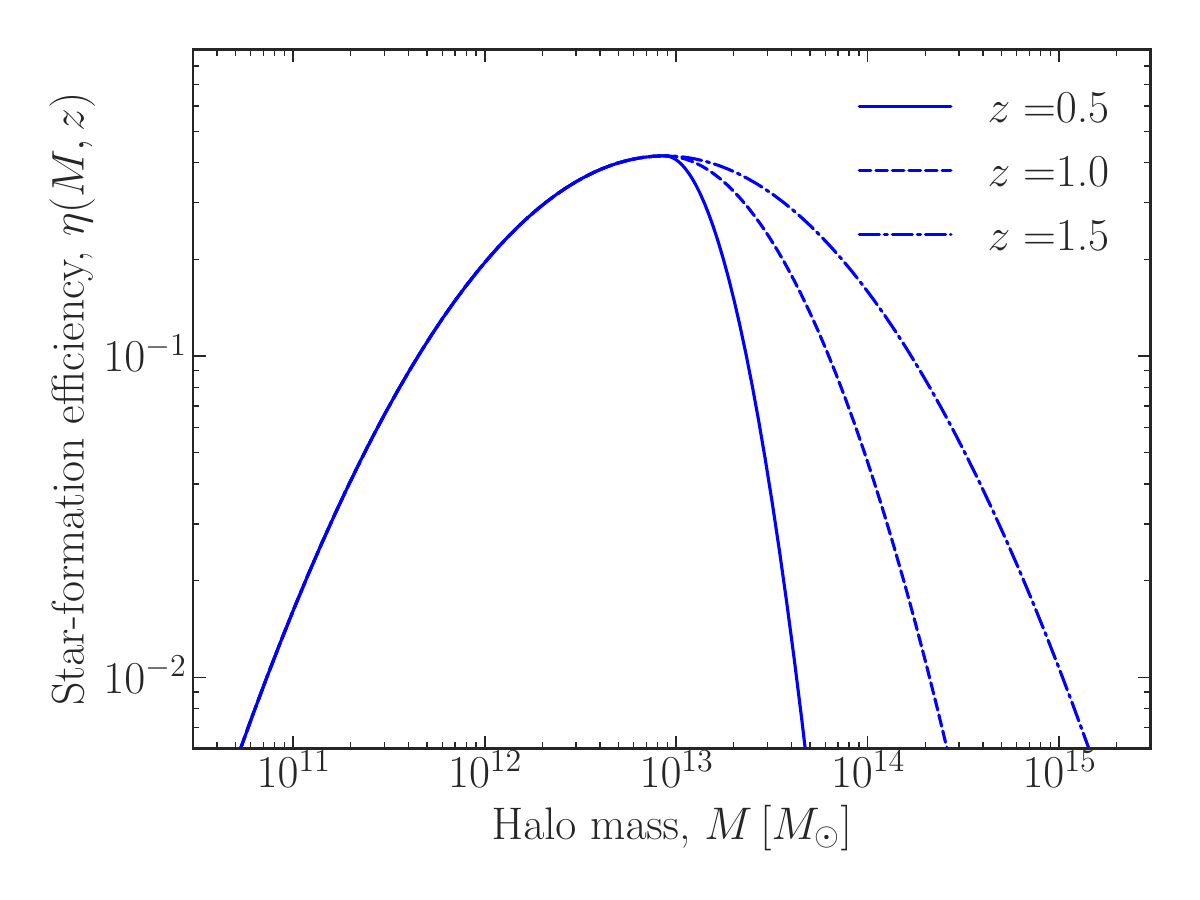}
\caption{Mass-dependent star-formation efficiency at various redshifts.}
\label{fig:eta}
\end{figure}

The star-formation efficiency peaks at a specific halo mass and declines for lighter and heavier haloes. This behavior is motivated by the effect of low gravitational potentials and supernova feedback at low masses, whereas the decrease at higher masses is due to an extended gas cooling time. In \cref{fig:eta}, we illustrate the star-formation efficiency in the \citetalias{maniyar2021simple} model as a function of halo mass at $z=0.5$, $1.0$, and $1.5$.
The peak efficiency here is constant as a function of time, and all time evaluations are incorporated into the extent of the high-mass tails of $\eta(M,z)$.

On the other hand, the contribution to the SFR from satellite galaxies is determined by the cumulative contribution of all sub-haloes associated with each parent halo of mass $M$, which we can write as
\begin{equation}
    \sfr_{\rm s}(M, z)=\int_{M_{\rm min}}^{M}\de M_{\rm sub}\,\frac{\de N_{\rm sub}}{\de M_{\rm sub}}\,\sfr_{\rm sat}(M_{\rm sub}, M, z)\,,
\end{equation}
with $\de N_{\rm sub}/\de M_{\rm sub}$ being the sub-halo mass function -- i.e., number of sub-haloes of mass \(M_{\rm sub}\) per halo -- calculated as in \citet{2010ApJ...724..878T}, and $\sfr_{\rm sat}(M_{\rm sub}, M, z)$ is the \sfr\ in a satellite galaxy with sub-halo mass $M_{\rm sub}$ within a parent halo of mass $M$. Here, $M_{\rm min}=10^5 M_\odot$ is the minimum subhalo mass. As in \citetalias{maniyar2021simple}, we model $\sfr_{\rm sat}$  as 
\begin{equation}
    \sfr_{\rm sat}(M_{\rm sub}, M, z)=
    \min\left[\sfr_{\rm c}(M_{\rm sub}, z), \frac{M_{\rm sub}}{M}\,\sfr_{\rm c}(M, z)\right]\,,
\end{equation}
which guarantees that the \sfr\ in any satellite never exceeds that of the central galaxy.

\begin{figure}[htbp!]
\centering
\includegraphics[angle=0,width=1.0\columnwidth]{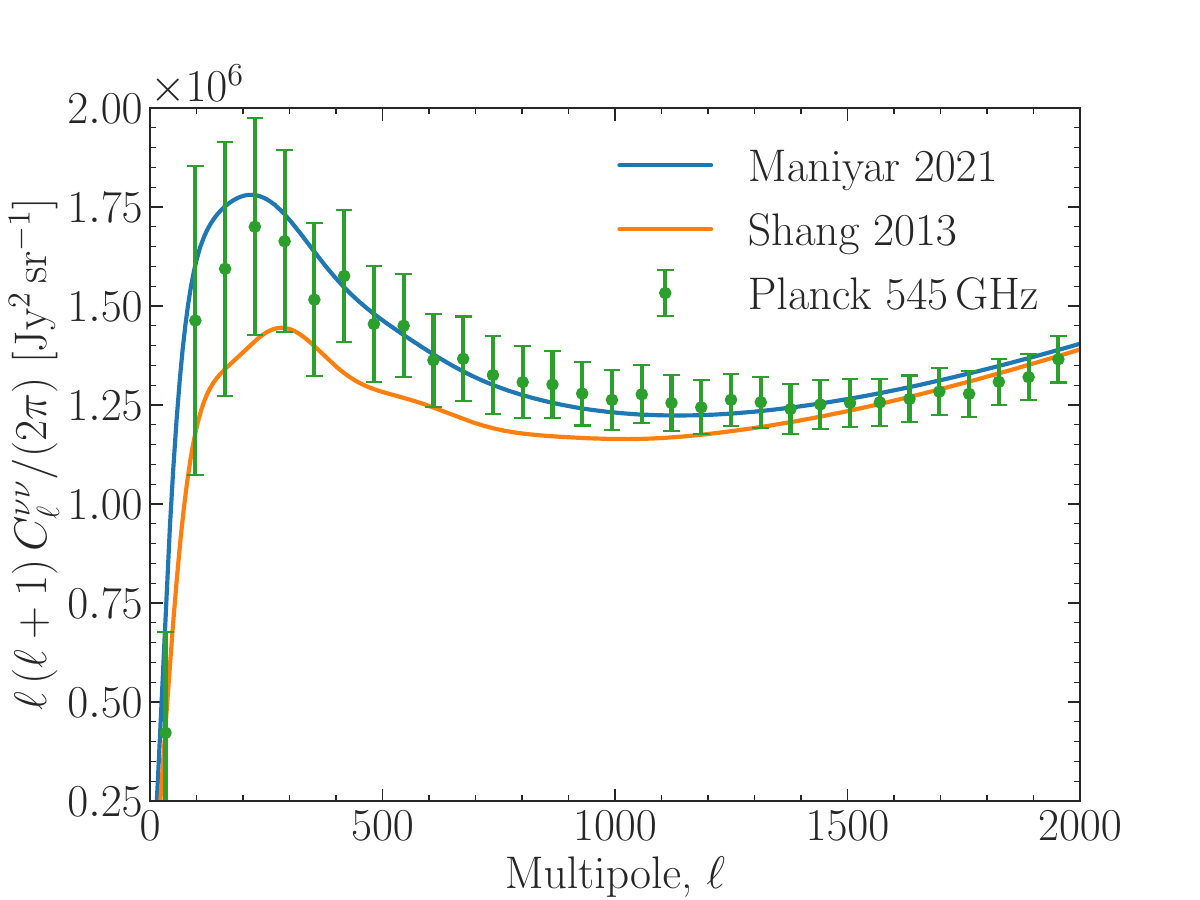}
\caption{Halo model power spectrum for the CIB auto-correlation. The green points with the corresponding error bars are the \Planck CIB data at $545\,\mathrm{GHz}$.}
\label{fig:cellcib}
\end{figure}

\Cref{fig:cellcib} displays the auto-power spectrum of the CIB. The blue curve represents the prediction from the \citetalias{maniyar2021simple} model, while the orange curve shows the prediction from the halo model of \citet{2012MNRAS.421.2832S}. The green points with associated uncertainties correspond to the \Planck data \citep{Lenz:2019ugy}. The two models differ primarily in their treatment of the galaxy-halo connection and the modeling of star formation within halos. The \citetalias{maniyar2021simple} model adopts a simplified, physically motivated prescription that links infrared luminosity to halo mass directly through an empirical star-formation efficiency; the \citet{2012MNRAS.421.2832S} model relies on a more standard halo occupation distribution framework in which infrared-emitting galaxies are populated into halos using halo occupation functions, with star formation encoded through the assumed luminosity-halo mass relation. Here, the shot-noise contribution included in the model curves is taken from [table] 3 of \citet{ade2011planck}.

The \citetalias{maniyar2021simple} halo model (blue curve) shows better agreement with the CIB auto-power spectrum data from \Planck (green curve) at large angular scales. This suggests that the \citetalias{maniyar2021simple} model more accurately captures the large-scale structure of the CIB than the most commonly used CIB halo model \citep{2012MNRAS.421.2832S}, owing to its refined treatment of the underlying halo distribution and star-formation processes.

\section{\label{tomo}Tomographic template fitting}
A key feature of projected galaxy clustering is its local nature: galaxies at redshift $z$ lie within structures at the same redshift. This allows for the direct measurement of the redshift evolution of astrophysical quantities by cross-correlating a specific projected tracer (e.g., CIB intensity maps) with galaxy samples in different redshift bins, a method commonly referred to as `tomography'. To do this, it is necessary to have some amount of redshift information for the galaxies in the catalog, so that we can slice their redshift distribution up into $i=1,\ldots, N_z$ redshift bins. This corresponds to having $n(z)=\sum_i n_i(z)$, which in turn allows us to define the galaxy clustering kernel of \cref{eq:delta} in each bin, such that $q_{\rm g}(z)=\sum_iq_{{\rm g},i}(z)$. We can then obtain a tomographic matrix for the power spectrum of galaxy clustering, $\smash{C^{\rm gg}_{ij}(\ell)}$, where $i=j$ represents auto-bin correlations and $i\ne j$ cross-bin correlations. 

Now, consider \cref{eq:cell,eq:large_scale,eq:qnu,eq:delta}, and assume that the radial dependence of $b_{\rm g}$, $\brsfrd$, and $q_\nu(z)$ varies slowly relative to the width of the galaxy kernel in a given redshift bin, $q_{{\rm g},i}(z)$. Then, the auto-correlation of galaxies and their cross-correlation with the CIB can be approximated using a template \citep[see also][]{2023ApJ...948....6T}, and we can write
\begin{align}
    C^{\rm gg}_{ij}(\ell)&\simeq b_{{\rm g},i}\,b_{{\rm g},j}\,T^{\rm gg}_{ij}(\ell)+\frac{\delta^{\rm K}_{ij}}{\bar{n}_i}\,,\label{eq:template_gg} \\
    C^{\rm g\nu}_i(\ell)&\simeq b_{{\rm g},i}\,\brsfrd_i\,T^{\rm g\nu}_i(\ell)\,,\label{eq:template_gnu}
\end{align}
where $b_{{\rm g},i}$ and $\brsfrd_i$ are calculated at the mean redshift of the $i$-th bin and $\delta^{\rm K}$ is the Kronecker-delta symbol. The power spectrum templates are defined as
\begin{equation}
    T^{u v}(\ell)\coloneqq\int^{z_{\rm hi}}_{z_{\rm lo}}\frac{c\,\de z}{H(z)}\,\frac{q_u(z)\,q_v(z)}{r^2(z)}\,P\left[k = \frac{\ell+1/2}{r(z)}, z\right]\,,
    \label{eq:muv}
\end{equation}
where $P(k,z)$ is the matter power spectrum (whose linear counterpart we have already introduced as $P_{\rm lin}$), while
\begin{equation}
    n_{u v}\coloneqq\int^{z_{\rm hi}}_{z_{\rm lo}}\frac{c\,\de z}{H(z)}\,\frac{q_u(z)\,q_v(z)}{r^2(z)}\,N_{U V}(z)\,,
    \label{eq:nuv}
\end{equation}
plays the role of a noise term for the projected spectra, sourced by the $N_{UV}$ term of \cref{eq:large_scale}. Under the assumption of a fixed cosmology and precise knowledge of the radial kernels, $T^{\rm gg}(\ell)$ and $T^{\rm g\nu}(\ell)$ can be pre-computed and treated as fixed templates. Thus, the galaxy auto-correlation provides constraints on the value of $b_{\rm g}$, which can then be used to constrain $\brsfrd$ at the redshift of the sample through the cross-correlation.

\begin{figure}[htbp!]
\centering
\includegraphics[angle=0,width=1.0\columnwidth]{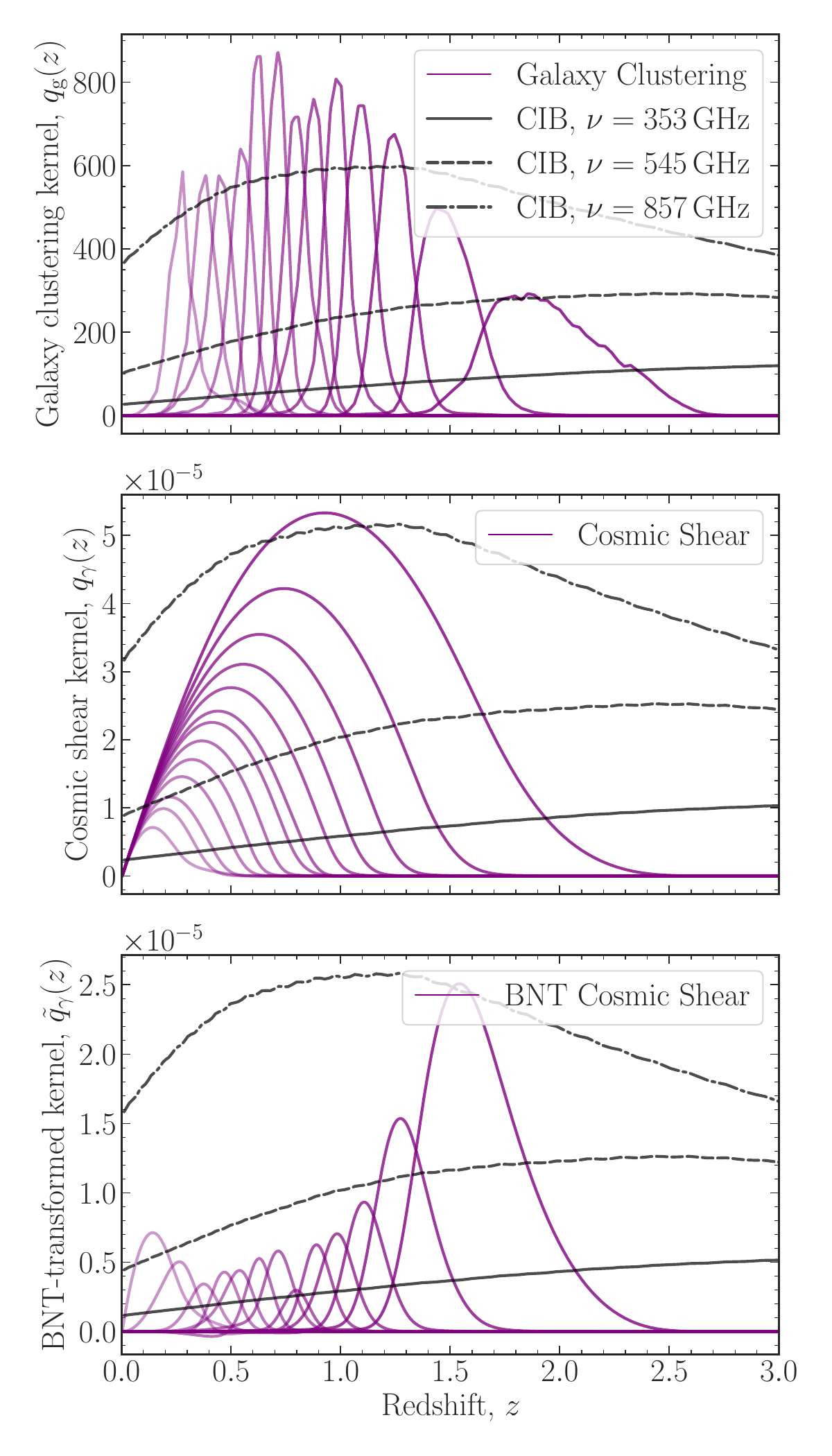}
\caption{Galaxy clustering kernels are shown in the upper panel, \( q_{\rm g}(z) \), corresponding to the 13 \Euclid\ photometric redshift bins. The middle panel shows the tomographic cosmic shear kernels, \( q_{\gamma,i}(z) \). The lower panel illustrates the BNT-transformed kernels, \( \tilde{q}_{\gamma,i}(z) \), which are more localized in redshift. CIB kernels are also shown in all panels.}
\label{fig:gamma}
\end{figure}

\Cref{fig:gamma} shows the radial kernels of the tracers used in this study. The CIB kernels are nearly constant across the 13 redshift bins of the \Euclid photometric galaxy catalog. This justifies our use of the approximation model of \cref{eq:template_gg,eq:template_gnu} to perform tomographic measurements of $\brsfrd$. However, to implement the tomographic template-fitting method for cross-correlating CIB and cosmic shear, the key requirement that the CIB kernel vary slowly relative to the cosmic shear kernel is not satisfied, as is evident from the integrated form of \cref{eq:qgamma}. To overcome this issue, we follow \citet{2022arXiv220603499C} and make use of the so-called BNT nulling technique, to localise lensing kernels in redshift. The BNT-transformed kernels, denoted with a tilde, are given by
\begin{equation}
    \tilde q_{\gamma,i}(z) = M_{ij}\,q_{\gamma,j}(z)\,,
\end{equation}
with the BNT matrix, $\tens{M}$, \citep[which we compute as in][]{taylor2021x} being redshift independent.
The result of this operation is shown in \cref{fig:gamma}. In the upper panel, we plot the weak lensing kernel for cosmic shear, while the lower panel shows the effect of the BNT transform, demonstrating the localization of the shear kernel. 

Thanks to the BNT transform, we can then apply the same template approach to the cross-correlation between cosmic shear and the CIB. Analogously to \cref{eq:template_gnu}, we can write
\begin{equation}
    \tilde C^{\epsilon\nu}_i(\ell)\simeq\brsfrd_i\,\tilde T^{\epsilon\nu}_i(\ell)+\tilde n_{\epsilon\nu}\,,\label{eq:template_ggamma}
\end{equation}
where both template and projected noise can be pre-computed and treated as fixed templates, as in \cref{eq:muv,eq:nuv}.

\section{\label{res}Results and discussion}
Here, we first present the forecast signals for the approach described in the previous section and then move on to the expected constraints on SFR history with our tomographic templates.

\subsection{Power spectra}
We computed all auto- and cross-power spectra using the publicly available Python package \texttt{pyccl} \citep{Chisari_2019},\footnote{\url{https://github.com/LSSTDESC/CCL}} that provides a consistent framework for modeling large-scale structure observables within a $\Lambda$CDM cosmology. Regarding cosmological parameters, we adopt those from \citet{EuclidSkyOverview}. For the halo-model quantities, we use the halo mass function of \citet{2008ApJ...688..709T} and compute the halo bias following \citet{2010ApJ...724..878T}. The concentration-mass relation is taken from \citet{2008MNRAS.390L..64D}, while the halo density model is described by the Navarro--Frenk--White profile \citep{navarro1996structure,navarro1997universal}.

We construct the tomographic galaxy auto-correlation power spectrum across the 13 photometric redshift bins, along with their cross-power spectrum with the three CIB maps (at \SI{353}{GHz}, \SI{545}{GHz}, and \SI{857}{GHz}) from \cite{Lenz:2019ugy}. These were extracted from \Planck data through a dedicated component separation processing using external HI data to trace and remove the Galactic dust emission. However, the cleaning pipeline of \citet{Lenz:2019ugy} applies a relatively aggressive Galactic mask to exclude the most contaminated regions of the sky. As a consequence, the overlap between the resulting CIB maps and the \Euclid survey footprint is reduced to approximately 17 percent of the sky, which limits the statistical power of the cross-correlation measurements.

To fully assess the constraining power that can be extracted from the available \Planck data, we additionally perform the same analysis using the GNILC-cleaned CIB maps available from the \Planck Legacy Archive~\citep{PlanckCIBremazeille}, focusing in particular on the \SI{545}{GHz} channel. Unlike the maps of \citet{Lenz:2019ugy}, the GNILC (Generalized Needlet Internal Linear Combination) approach separates Galactic foregrounds and CIB emission directly from the multifrequency \Planck observations by exploiting their distinct statistical properties across angular scales and frequencies. This methodology enables a larger fraction of the sky to be retained after foreground cleaning, increasing the overlap with the \Euclid survey footprint $f_{\rm sky}$ from $17\%$ to $27\%$. We therefore use the GNILC maps as a complementary dataset to quantify the impact of sky coverage on achievable parameter constraints and to assess the effect of the choice of CIB cleaning procedure on our results. We emphasize that the CIB maps enter our analysis only through the construction of the covariance matrix, while the fiducial angular power spectra employed in the Fisher forecasts are computed entirely from the theoretical model. As a result, differences between the \citet{Lenz:2019ugy} and GNILC products primarily affect the forecasts through their statistical characteristics, including the effective sky fraction and noise properties.

As discussed, we impose a default low-$\ell$ cut of $\ell_{\rm min} = 100$ to account for potential systematics in the tomographic analysis. In addition to the validity of the Limber approximation, this choice is motivated by the suppression of large-scale power in CIB maps resulting from the removal of Galactic contamination across different survey patches, as reported by \citet{Lenz:2019ugy}. On small scales, instead, we apply a redshift-bin-dependent upper multipole cut at $\ell_{\max,i} = k_{\max}\,r(\bar z_i) - 1/2$, where $r(\bar z_i)$ denotes the radial comoving distance to the mean redshift of each galaxy sample. While we set $k_{\max} = 0.15\,\mathrm{Mpc}^{-1}$ for our analysis, we also tested a more optimistic scenario with $k_{\max} = 0.3\,\mathrm{Mpc}^{-1}$. This allows us to estimate the potential improvement from including additional small-scale modes. 

\begin{figure*}[htbp!]
\centering
\includegraphics[angle=0,width=1.0\textwidth]{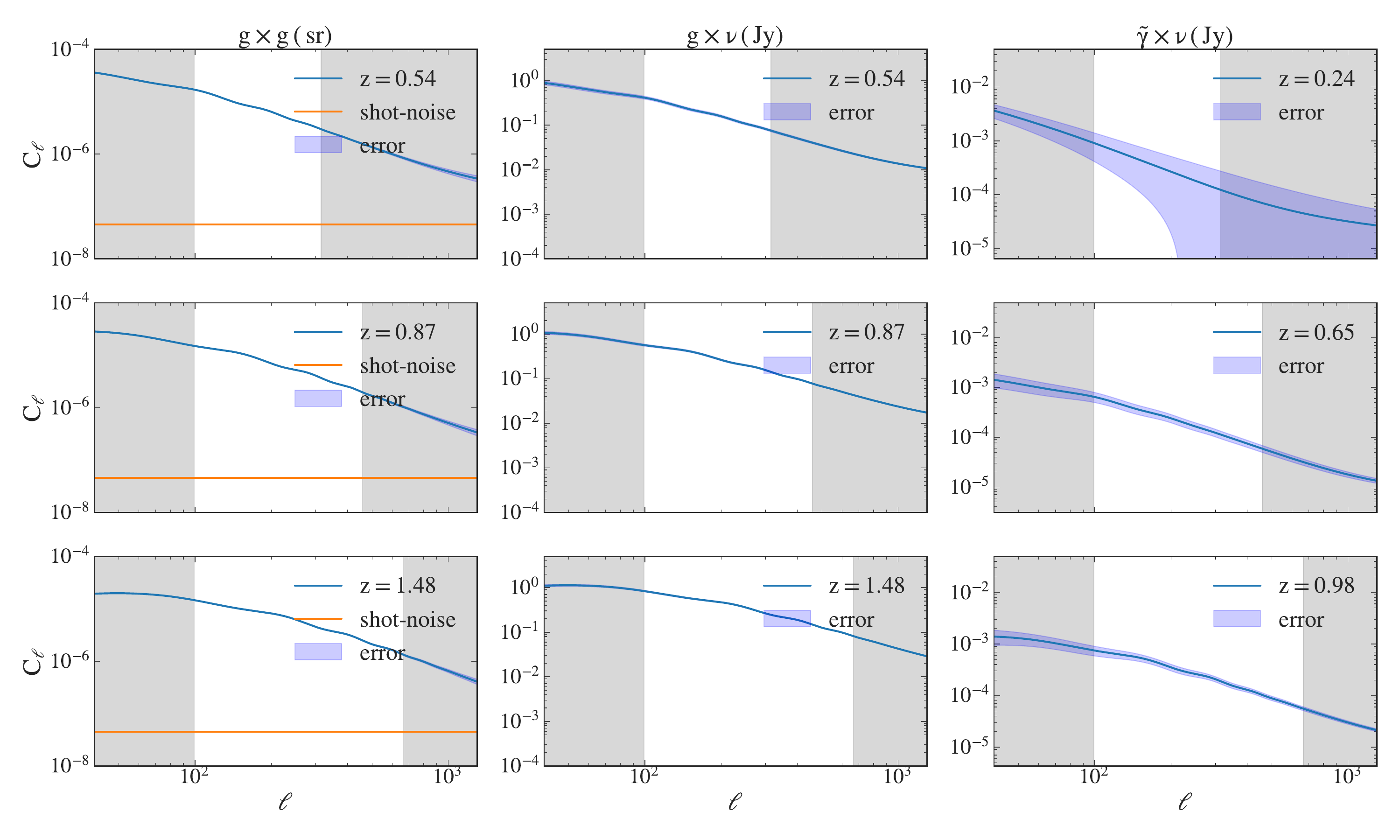}
\caption{Galaxy auto-spectra (left column), galaxy--CIB cross-spectra (middle column), and shear--CIB cross-spectra (right column). CIB observations refer to the \SI{545}{GHz} \Planck frequency channel, and for illustrative purposes we show only the 4th, 8th, and 12th tomographic redshift bins (see \cref{fig:zker,fig:gamma}). Each panel shows theoretical predictions and their corresponding $\pm1\,\sigma$ uncertainties (blue curve with blue-shaded area), while the orange line denotes the Poissonian shot-noise term for galaxy clustering. The light-grey vertical regions mark scales cut out of the analyses.}
\label{fig:ps}
\end{figure*}

\Cref{fig:ps} shows the predictions of all the quantities relevant to our analysis for a few example redshift bins: galaxy auto-correlation power spectra (left panels); galaxy--CIB cross-correlation spectra (central panels); and BNT-transformed shear--CIB cross-spectra (right panels). 
Shaded areas around the blue curves mark the $\pm1\,\sigma$ uncertainty (see \cref{cova}), whereas the orange lines in the left panels are for the galaxy shot-noise term, $n_{\rm gg}$. The cross-noise is assumed to be negligible, as is usually the case, such that \ $n_{\rm g\nu}=n_{\epsilon\nu}\equiv0$.

\subsection{\label{cova}Information matrix}
To quantify the cross-correlation between the CIB and \Euclid probes and to estimate the constraining power of the tomographic template method on \brsfrd, we perform an information-matrix analysis.

To begin with, we first need an ansatz for the covariance matrix of the measured spectra, which is the sum of signal and noise, as in \cref{eq:template_gg,eq:template_gnu,eq:template_ggamma}. For simplicity, we assume a Gaussian covariance, which takes the form
\begin{equation}
\cov[C^{uv}_{ij}(\ell), C^{uv}_{mn}(\ell')] = \frac{C^{uu}_{im}(\ell)\,C^{vv}_{jn}(\ell) + C^{uv}_{in}(\ell)\,C^{uv}_{jm}(\ell)}{(2\,\ell + 1)\,\Delta\ell\,f_{\rm sky}}\,\delta_{\ell \ell'}\,,\label{eq:covariance}
\end{equation}
where $\Delta\ell$ is the width of the multipole bin and $f_{\rm sky}$ is the fraction of the sky area covered by both \Euclid and \Planck. Consistent with the \Planck CIB analysis \citep{ade2014planck}, we adopt $\Delta\ell=64$. The non-vanishing noise terms for the \Euclid probes entering the covariance are
\begin{equation}
n_{\rm gg}=1/\bar n\,, \qquad n_{\epsilon\epsilon}=\sigma_\epsilon^2/\bar n\,,
\end{equation}
with $\bar n=1.87\,\mathrm{arcmin}^{-2}$ being the surface density of the equi-populated bins adopted here, $\sigma_\epsilon^2$ denotes the variance of measured galaxy ellipticities (here, we adopt $\sigma_\epsilon=0.3$). For $n_{\nu\nu}$, the shot-noise term of the auto-power spectrum of CIB, it is included in the noise power spectrum calculated from the \Planck observed datasets.

For an $\ell$-diagonal covariance matrix as in \cref{eq:covariance}, the information matrix is
\begin{equation}
    I_{\alpha\beta} = \sum_\ell \sum_{ijmn} \frac{\partial C^{uv}_{ij}(\ell)}{\partial p_{\alpha}}\,\cov^{-1}[C^{uv}_{ij}(\ell), C^{uv}_{mn}(\ell)]\,\frac{\partial C^{uv}_{mn}(\ell)}{\partial p_{\beta}}\,,
    \label{eq:fisher}
\end{equation}
where $p_\alpha$ is the $\alpha$-th cosmological parameter. In this study, the free parameters are $b_{{\rm g},i}$ and $\brsfrd_i$.

\begin{figure*}[htbp!]
\centering
\includegraphics[angle=0,width=0.8\textwidth]{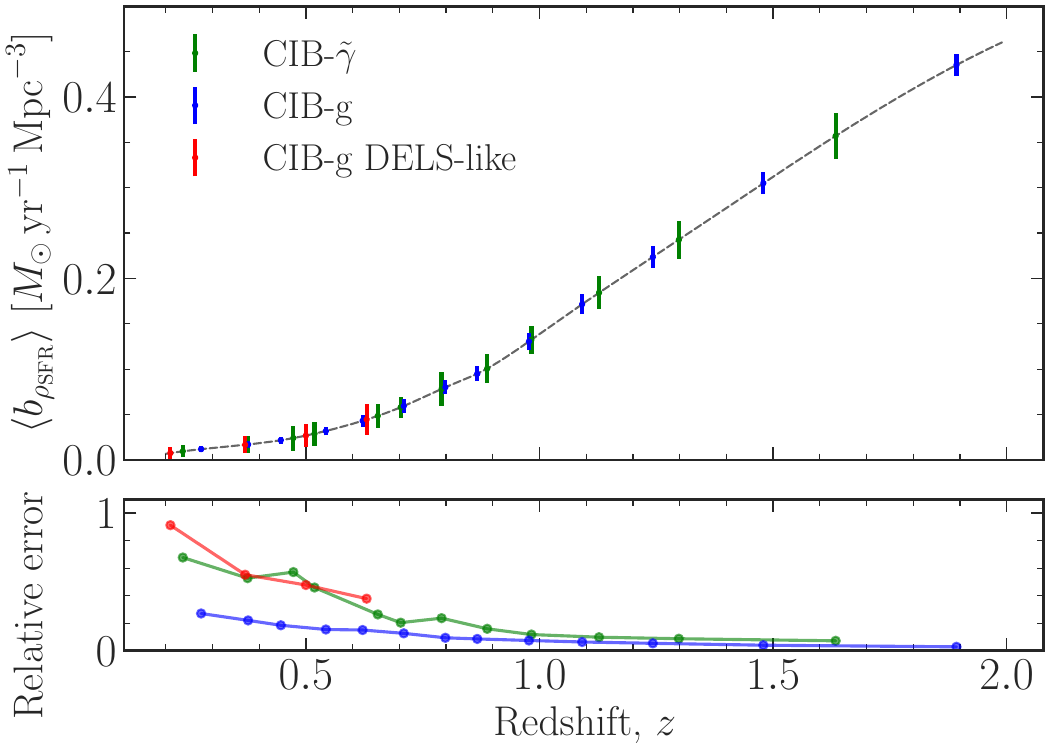}
\caption{Forecast marginal $1\,\sigma$ constraints on $\brsfrd$ from the cross-correlation between the \SI{545}{GHz} CIB map of \citet{Lenz:2019ugy} and BNT-transformed shear (green) or photometric galaxy number counts (blue). The dashed curve represents the theoretical prediction. For comparison, we also show a forecast re-analysis of the galaxy--CIB DELS-like cross-spectra, which are used in \citet{Jego:2022eqo}.The relative error is defined as $\sigma_i / \brsfrd_i$, where $\sigma$ denotes the $1\sigma$ uncertainty.}
\label{fig:result}
\end{figure*}

By combining \cref{eq:fisher,eq:template_gg,eq:template_ggamma,eq:template_gnu}, we obtain the information matrix for the template-fitting method.
Then, given the Cram\'er--Rao bound \citep{rao1992information,cramer1999mathematical}, we can obtain a lower limit on the variance of a parameter $p_{\alpha}$ from the corresponding diagonal element of the inverse Fisher information matrix, namely
\begin{equation}
\label{sigma}
\sigma^2(p_{\alpha}) = \left(\tens I^{-1} \right)_{\alpha\alpha}\,.
\end{equation}
Thus, we can estimate the constraining power of the template-fitting method. In \cref{fig:result}, we show the forecast constraints for the two cross-correlation tomographic template-fitting methods in this study. We present only the \SI{545}{GHz} CIB measurement as an example. Under the forecast settings adopted in this work, all three \Planck channels produce comparable constraining power. 
In \cref{fig:result}, the galaxy--CIB cross-power spectrum is plotted in blue, and the error bars represent the $1\,\sigma$ marginal uncertainties. The shear--CIB cross-power spectrum is shown in green, and the effective mean redshift for each bin differs from that of the galaxy because the weighting functions differ between the galaxy and cosmic shear kernels. The curve represents the theoretical prediction of $\brsfrd$.

We also seek to compare these estimates with current results based on the cross-spectrum between the DELS and the \citet{Lenz:2019ugy} \Planck CIB observations. However, a direct comparison with \citet{Jego:2022eqo} would not be fair, since their analysis is based on real DELS and \citet{Lenz:2019ugy} \Planck CIB maps. As such, their analysis is more realistic and includes systematic effects that we do not take into account. Hence, we perform a forecast analysis of DELS--CIB and rerun the forecast pipeline, replacing our data set with the photometric sub-sample of DELS selected by \citet{hang2021galaxy}.  Our results are plotted in \cref{fig:result} in red. Regarding the constraining power, the \Euclid galaxy--CIB data set has approximately two times tighter constraints than the DELS--CIB data set, 
while the \Euclid shear--CIB data set is about the same level and covers a much larger redshift interval. Unlike~\citet{Jego:2022eqo}, we do not include eBOSS in our DELS--CIB forecast analysis because it includes both galaxy and quasar samples, whereas our goal is to maintain a consistent comparison with \Euclid, which considers only galaxy samples. In addition, the infrared properties of quasars are not comparable to those of the DELS galaxy population, since their emission is largely dominated by AGN-related processes rather than star formation.

We also test a more optimistic small-scale scenario by increasing $k_{\max} = 0.3\,\mathrm{Mpc}^{-1}$ in the \Euclid forecasts. Including these additional modes captures a larger fraction of the galaxy clustering signal and leads to a modest average improvement of approximately $17\,\%$ in the constraints on $\brsfrd$. Although this illustrates the potential gains from pushing to smaller scales, extending the analysis beyond $k_{\max} = 0.15\,\mathrm{Mpc}^{-1}$ requires caution. In particular, nonlinear effects and the limitations of the halo occupation distribution model become increasingly important on these scales. As discussed by \citet{mead2021including}, non-linearities primarily affect the scale and smoothness of the transition between the 1-halo and 2-halo regimes. A consistent treatment of this transition requires detailed modeling of nonlinear mode coupling, halo exclusion, and scale-dependent bias, as well as issues related to SFR modeling, which is not straightforward to implement. We therefore do not attempt to include these effects here, as that would go beyond the scope of the present forecast analysis.

We further perform a Fisher information matrix analysis of the parameters of the halo model introduced in \cref{CIB} by projecting the Fisher matrix, initially computed for parameters $b_{\rm g}$ and $\brsfrd$, onto the new set of parameters (see \Cref{apdx:A} for details of the derivation). Applying \cref{sigma} to the Fisher matrix of the halo model parameters, we obtain the projected uncertainties $\sigma$, which characterize the constraining power of our analysis.

\begin{figure*}[htbp!]
\centering
\includegraphics[angle=0,width=0.8\textwidth]{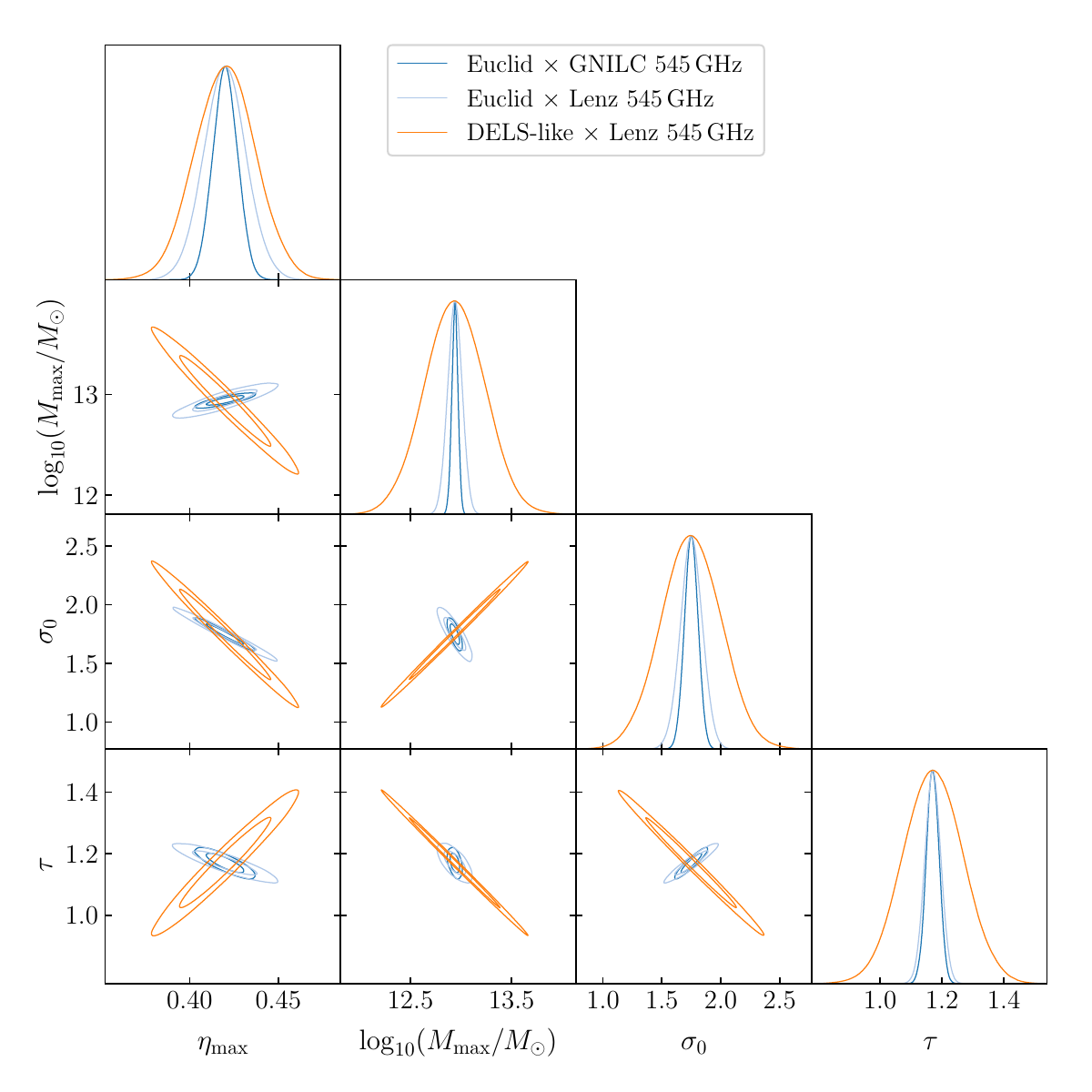}
\caption{Forecasts of posterior distributions for the HOD parameters, with the cross-correlation of \Euclid photometric galaxy clustering with CIB data at \SI{545}{GHz} from \citet{Lenz:2019ugy} and GNILC-cleaned \Planck, shown in light blue and dark blue. We also show DELS-like data set results in orange for comparison.}
\label{fig:corner}
\end{figure*}

We show the posterior distributions for the HOD parameters that describe the CIB emission in \cref{fig:corner} and their corresponding $1\,\sigma$ constraints in \cref{tab:params}, both of which were derived from Fisher information matrix analyses. The parameters shown are the characteristic halo mass $\log_{10} (M_{\rm max}/M_{\odot})$, the maximum star-formation-efficiency $\eta_{\rm max}$, the maximum width of the star-formation-efficiency function $\sigma_{0}$, and the characteristic timescale $\tau$. Visible degeneracies, especially between $\eta_{\rm max}$ and $\log_{10} (M_{\rm max}/M_{\odot})$, reflect the trade-off between the halo mass scale and efficiency in shaping the CIB signal. Significant improvements are obtained with \Euclid galaxy clustering compared to DELS. For both \Planck\ foreground-cleaning pipelines considered in this analysis, \Euclid galaxy clustering significantly tightens the constraints on all HOD parameters, notably reducing the $\eta_{\rm max}$--$\log_{10} (M_{\rm max}/M_{\odot})$  degeneracy and improving the precision on $\sigma_{0}$ and $\tau$ by roughly $150\,\%$. 

To assess the impact of the adopted CIB cleaning procedure, we compare the resulting constraints obtained using the \citet{Lenz:2019ugy} maps with those derived from the GNILC-cleaned \Planck products in \cref{fig:corner}. The GNILC-based forecasts yield consistently tighter constraints than those obtained with the \citet{Lenz:2019ugy} maps. Since the fiducial signal is identical in both cases and the CIB maps enter the analysis only through the covariance matrix, this improvement is mainly driven by the increased overlap with the \Euclid\ footprint $(f_{\rm sky}\simeq 0.27$ versus $0.17$). The resulting reduction in statistical uncertainties is particularly evident for $\eta_{\rm max}$, a parameter that predominantly controls the overall amplitude of the CIB signal, and is therefore especially sensitive to improvements in the signal-to-noise ratio.

To assess the impact of including smaller scales, we repeat the analysis with a higher $k_{\rm max} = 0.3~\mathrm{Mpc}^{-1}$. This leads to an improvement of approximately $10\,\%$ in the constraints of the HOD parameters and partially alleviates the $\eta_{\rm max}$--$\log_{10} (M_{\rm max}/M_{\odot})$ degeneracy. However, non-linearities and the breakdown of the HOD model at these scales require additional treatment to ensure robust estimates.

\begin{table}[htbp]
    \centering
    \caption{Fiducial parameters and $1\,\sigma$ uncertainties.}
    \label{tab:params}
    \begin{tabular}{lrr}
        \hline
        \hline
        Parameter & Fiducial value & Uncertainty \\
        \midrule
        $\log_{10} (M_{\rm max}/M_{\odot})$ & 12.94 & 0.15 \\
        $\eta_{\rm max}$        & 0.420 & 0.015 \\
        $\sigma_{0}$          & 1.75  & 0.10 \\
        $\tau$                  & 1.17  & 0.03 \\
        \hline
    \end{tabular}
    \tablefoot{ Fiducial parameter values and corresponding $1\,\sigma$ uncertainties for the cross-spectrum between \Euclid galaxy clustering and the \Planck CIB $545\,\mathrm{GHz}$.
    }
\end{table}

\section{Conclusions}
\label{sc:conc}
The main purpose of this paper is to demonstrate the potential of combining {\Euclid}’s photometric galaxy clustering and weak lensing data with CIB measurements to improve our understanding of star-formation processes over time. We have presented a tomographic analysis of the CIB through forecast cross-correlations with 13 photometric galaxy redshift bins from the \Euclid wide survey, with redshifts ranging up to $z\simeq2$. Following previous analyses in the literature, we exploit photometric galaxy number counts to trace star-formation history. Our analysis enables us to directly constrain the evolution of the bias-weighted SFRD $\brsfrd$ as a function of redshift. Furthermore, for the first time, we propose using the weak lensing effect of cosmic shear for this purpose, employing the BNT nulling technique, which allows us to localize the integrated lensing signal as a function of redshift.

We perform a three-step analysis. First, we construct tomographic templates for each power spectrum. In the second step, from these we compute power spectra and their associated variances, including galaxy auto-correlation power spectra, galaxy--CIB cross-spectra, and shear--CIB cross-spectra. 
Finally, we perform an information matrix analysis to assess the constraining power of the data set on the model parameters.

Our analysis yields a constraint on the target parameter that is approximately twice as tight as that obtained by applying the same methodology to the DELS photometric galaxy sample used in a previous study \citep{Jego:2022eqo}. In addition, we perform a cosmic shear analysis, which provides independent constraints of comparable precision to those derived from the galaxy--CIB cross-correlation analysis of the DELS data set. With the photometric galaxy samples from \Euclid, we obtain a data set with a signal-to-noise ratio ($\mathrm{S/N}$) of approximately 125 under ideal conditions. When including the cosmic shear component, which contributes an $\mathrm{S/N}$ of 67, a joint analysis combining both probes increases the total $\mathrm{S/N}$ to 132 (assuming no cross-covariance). For comparison, applying our pipeline to the DELS-like specification used in \citet{Jego:2022eqo} yields a total $\mathrm{S/N}$ of 58. With a noticeably lower $\mathrm{S/N}$, as shown in \cref{fig:corner}, the DELS--CIB case exhibits broader posteriors for all parameters, with a pronounced degeneracy between $\eta_{\rm max}$ and $\log_{10} (M_{\rm max}/M_{\odot})$. In contrast, the \Euclid forecast significantly tightens the HOD constraints, reducing the $\eta_{\rm max}$--$\log_{10} (M_{\rm max}/M_{\odot})$ degeneracy and improving the precision on $\sigma_{0}$ and $\tau$ by roughly $150\,\%$. These results demonstrate that, with deeper and more tomographically resolved galaxy samples, \Euclid will further enhance constraints on and understanding of HOD models.

With the advent of experiments like \Euclid, we will be able to trace the evolution of star formation with unprecedented precision and depth in cosmic time. In addition, high-resolution measurements from forthcoming experiments such as Cerro Chajnantor Atacama Telescope \citep[CCAT,][]{stacey2018ccat} and the Simons Observatory \citep{ade2019simons} are expected to significantly enhance the quality, sensitivity, and spectral coverage of CIB observations. We have tested forecasts using CCAT noise, finding that differences appear primarily at very high multipoles, where the superior angular resolution of CCAT provides an advantage on small scales (see also \citealt{mccarthy2021improving}). However, these scales are effectively cut off by our current $k_{\rm max}=0.15\,{\rm Mpc}^{-1}$ limit, so CCAT will not have a significant impact on our fiducial analysis. However, if the HOD model could be better constrained on small scales, allowing us to extend the applicable range to $k_{\rm max} = 0.3\,\mathrm{Mpc}^{-1}$, we forecast an approximate $5\,\%$ increase in the total $\mathrm{S/N}$ for the joint analysis compared to that of the \Planck CIB data, since the CCAT noise is substantially lower for $\ell \gtrsim 1000$. Thus, these improved measurements of the CIB will further enhance our ability to study the star-formation history.

\begin{acknowledgements}
\AckEC
SC acknowledges support from the Italian Ministry of University and Research (\textsc{mur}), PRIN 2022 `EXSKALIBUR – Euclid-Cross-SKA: Likelihood Inference Building for Universe's Research', Grant No.\ 20222BBYB9, CUP D53D2300252 0006, and from the European Union -- Next Generation EU. MM acknowledges support from the INFN InDark project.
\end{acknowledgements}

\bibliography{Euclid, Q1, my} 

\begin{appendix}
\onecolumn 
  
\section{\label{apdx:A}Jacobian matrix}

In this appendix, we derive the Jacobian matrix that projects the Fisher matrix from the template-fit parameters to the halo occupation distribution (HOD) parameters introduced in \cref{probe}. The Fisher analysis is initially performed in terms of the galaxy bias \(b_{\rm g}\) and the bias-weighted star-formation-rate density \(\brsfrd\) in each redshift bin, which are the quantities directly constrained by the tomographic template fitting procedure. Constraints on the HOD parameters \(\{\eta_{\max}, \log_{10}(M_{\max}/M_\odot), \sigma_0, \tau\}\) are then obtained through a Jacobian transformation, whose derivation is presented below.

The Jacobian matrix associated with the transformation from the template-fit parameter set,
$\vec p=\{b_{{\rm g},i=1\ldots N_z},\brsfrd_{i=1\ldots N_z}\}$,
to the HOD parameter set,
$\vec\vartheta=\{\eta_{\max},\log_{10}(M_{\max}/M_\odot),\sigma_0,\tau\}$\,,
is defined as
\begin{equation}
    \tens J \equiv \frac{\partial \vec p}{\partial \vec\vartheta}\,,
\end{equation}
whose transpose can be written as
\begin{equation}
    \tens J^{\sf T}=
    \begin{pmatrix}
    0 & \ldots & 0 & \pd\brsfrd_1/\pd\eta_{\max} & \pd\brsfrd_1/\pd\log_{10}(M_{\max}/M_{\odot}) & \pd\brsfrd_1/\pd\sigma_{0} & \pd\brsfrd_1/\pd\tau \\
    \vdots & \ddots & \vdots & \vdots & \vdots & \vdots & \vdots \\
    0 & \ldots & 0 & \pd\brsfrd_{N_z}/\pd\eta_{\max} & \pd\brsfrd_{N_z}/\pd\log_{10}(M_{\max}/M_{\odot}) & \pd\brsfrd_{N_z}/\pd\sigma_{0} & \pd\brsfrd_{N_z}/\pd\tau \\
    \end{pmatrix}\,.
\end{equation}
Hence, the new information matrix on \(\vec\vartheta\) is related to the old one on \(\vec p\) (Eq.~\ref{eq:fisher}) via
\begin{equation}
    \tens I^{(\vec\vartheta)}=\tens J^{\sf T}\,\tens I^{(\vec p)}\,\tens J\,.
\end{equation}
Now, it is easy to see from \cref{eq:brsfr,eq:sfr,eq:sfr_c,eq:eta,eq:sigma_M} that, at any given redshift,
\begin{equation}
    \frac{\pd\brsfr}{\pd\vec\vartheta}=\int\de M \, n_{\rm h}(M) \, b_{\rm h}(M) \, \frac{\pd\sfr(M)}{\pd\vec\vartheta}    \,,
\end{equation}
with (omitting dependencies on both \(M\) and \(z\))
\begin{equation}
    \frac{\pd\sfr}{\pd\vec\vartheta}=\frac{\pd\eta}{\pd\vec\vartheta}\,{\rm BAR}+ \int_{M_{\rm min}}^{M}\de M_{\rm sub}\,\frac{\de N_{\rm sub}}{\de M_{\rm sub}}\,\frac{\pd\sfr_{\rm sat}}{\pd\vec\vartheta}(M_{\rm sub})\,,
\end{equation}
and
\begin{equation}
    \frac{\pd\sfr_{\rm sat}}{\pd\vec\vartheta}(M_{\rm sub},M)=\min\left[\frac{\pd\eta}{\pd\vec\vartheta}(M_{\rm sub})\,{\rm BAR}(M_{\rm sub}), \frac{\pd\eta}{\pd\vec\vartheta}(M)\,\frac{M_{\rm sub}}{M}\,{\rm BAR}(M)\right]\,,
\end{equation}
the last ingredients being
\begin{align}
    \frac{\pd\eta(M,z)}{\pd\eta_{\max}} &= \frac{\eta(M,z)}{\eta_{\max}} \,, \\
    \frac{\pd\eta}{\pd\log_{10} (M_{\rm max}/M_{\odot})} &= \frac{\eta(M,z)}{\left[ \sigma_M(M,z) \right]^2}\,\log_{10}\left(\frac M{M_{\max}}\right) \,, \\
    \frac{\partial \eta(M,z)}{\partial \sigma_{M,0}} &= \eta(M,z) \times \frac{[\log_{10} (M_{\max}/M)]^2}{\left[ \sigma_M(M,z) \right]^3}\,,\\
    \frac{\pd\eta(M,z)}{\pd\tau} &= \frac{\pd\eta(M,z)}{\partial \sigma_{M,0}}\,{\rm max}(0,z-z_{\rm c}) \,.
\end{align}
\end{appendix}
\label{LastPage}
\end{document}